\pdfoutput=1

\documentclass[11pt]{article}

\usepackage{acl2023}
\usepackage{times}
\usepackage{latexsym}
\usepackage[T1]{fontenc}
\usepackage[utf8]{inputenc}
\usepackage{microtype}
\usepackage{inconsolata}
\usepackage{epigraph}
\usepackage{longtable}
\usepackage{booktabs}
\usepackage{graphicx}

\title{Fingerprinting web servers through Transformer-encoded HTTP response headers}

\author{Patrick Darwinkel \\
\href{mailto:installgentoo@darwinkel.net}{\nolinkurl{<installgentoo@darwinkel.net>}}}

\begin{document}
	\maketitle
	\begin{abstract}
	We explored leveraging state-of-the-art deep learning, big data, and natural language processing to enhance the detection of vulnerable web server versions. Focusing on improving accuracy and specificity over rule-based systems, we conducted experiments by sending various ambiguous and non-standard HTTP requests to 4.77 million domains and capturing HTTP response status lines. We represented these status lines through training a BPE tokenizer and RoBERTa encoder for unsupervised masked language modeling. We then dimensionality reduced and concatenated encoded response lines to represent each domain's web server. A Random Forest and multilayer perceptron (MLP) classified these web servers, and achieved 0.94 and 0.96 macro F1-score, respectively, on detecting the five most popular origin web servers. The MLP achieved a weighted F1-score of 0.55 on classifying 347 major type and minor version pairs. Analysis indicates that our test cases are meaningful discriminants of web server types. Our approach demonstrates promise as a powerful and flexible alternative to rule-based systems.
	\end{abstract}

	\section{Introduction}

    \epigraph{Note: Revealing the specific software version of the server may
    allow the server machine to become more vulnerable to attacks
    against software that is known to contain security holes. Server
    implementers are encouraged to make this field a configurable
    option.}{RFC 2068 - HTTP/1.1 \\ \citet{rfc2068}}

    In the field of cybersecurity, it is well-established that having knowledge of software vendor and version
    is essential in locating vulnerabilities. Many public vulnerability databases exist, which
    contain information about which specific software version is vulnerable to which exploit. As such, systems that are not frequently updated are
    at risk of being vulnerable to attacks.
    This was common knowledge back in 1997, and the HTTP/1.1 protocol specification even contains a warning that
    the \texttt{Server:} header - which explicitly reveals this version information - should be implemented cautiously.
    System administrators should thus be careful to reveal as little of this information as possible to the public.
    Over 25 years later, a typical response to an HTTP request divulges this information by
    default\footnote{Some examples of ordinary HTTP response headers can be found in Figures
    \ref{figure:apache_example},        \ref{figure:nginx_example}, and \ref{figure:lighthttpd_example} in appendix \ref{sec:responseexamples}.},
    even though it is perfectly feasible to configure web server software to hide or falsify the
    \texttt{Server:} header.

    Software fingerprinting is essentially the process of attempting to determine the type and version
    of software based purely on its input and output behaviour.
    One of the first academic papers about software fingerprinting specifically in the context of web servers was published by
    \citet{lee2002detecting}.
    They demonstrated that hiding the \texttt{Server:} header is insufficient to counter all attempts at fingerprinting.
    The essence of their paper is that fingerprinting is possible because each software implementation of the HTTP
    protocol handles certain requests differently. \citeauthor{lee2002detecting} argue that this is because the specification leaves room for lexical, syntactic, and semantic ambiguity. This can be exploited by crafting several ambiguous HTTP requests, and comparing how
    different implementations respond to them.

    To the best of our knowledge, the first machine learning approach to web server fingerprinting was devised by \citet{book2013automated}.
    In order to study several distinguishing features of the World Wide Web, they trained a classifier to determine the confidence of a certain
    web server being of a certain type.
    They found that this was possible, although their methodology was not powerful enough to distinguish individual versions.

    There is very little public academic research on the topic of web server
    fingerprinting (as noted by both \citeauthor{book2013automated} in \citeyear{book2013automated} and \citeauthor{alhunting} in \citeyear{alhunting}), considering how fundamental the
    Hypertext Transfer Protocol is to almost all modern infrastructure. As a universal request-response
    protocol operating at the application layer of the Internet Protocol, it is not only used by web browsers
    to retrieve web pages. Its practical applications range from
    smart fridges submitting analytics and usage data to their original manufacturer,
    to self-driving vehicles requesting firmware updates.
    As of writing, classical rule-based systems\footnote{Such as \texttt{nmap}, \texttt{httprecon}, \texttt{httprint}, \texttt{hmap}, and \texttt{nikto}.}
    are still the standard tools for web server fingerprinting. These expert systems have some fundamental limitations as they
    do not scale well to increasingly large possible combinations of input,
    require labour-intensive feature engineering,
    and have no efficient strategy to rank similar, but not identical input. Although these existing systems still work well, we hypothesize that 20 years
    after the publication of the first research paper and the development of the first fingerprinting tool, it should
    be possible to develop a system that outperforms them in terms of accuracy and specificity.

    As such, the overall aim of this study is to investigate the feasibility of using state-of-the-art deep learning,
    big data, and natural language processing techniques to improve the ability to detect vulnerable web server versions,
    as compared to existing rule-based systems.
    As we will further outline in section \ref{section2}, there is grounding in academic literature that
    indicates that this is a promising, novel approach.

    First, we will break down our goal into concrete research questions, and explain how our work improves upon the experiment by \citet{book2013automated}.
    In section \ref{section2}, we will further discuss relevant aspects of web server fingerprinting, anomaly detection in the HTTP domain, and document embedding and representation.
    Section \ref{section3} continues by describing the source of our data and section \ref{section4} describes how we intend to use it to answer our research questions.
    We then discuss our findings in section \ref{section5}. We end this work with section \ref{section6}, in which we answer our research questions,
    and describe the limitations and potential for future work.

    \subsection{Research questions}

    \subsubsection*{Can we improve upon the accuracy of server type prediction set by \citet{book2013automated}?}

    We follow some of the suggestions that \citet{book2013automated} give for future work, namely increasing the amount of test cases, increasing the
    sample size, and having more data per sample (i.e., more than just a status response code for each test case).

    \subsubsection*{Can we build a classifier that can discriminate specific server versions?}

    The experiment by \citet{book2013automated} could not meaningfully discriminate between specific server software versions.
    The two challenges that arise are the lack of predictive features and the sparsity and variability of target classes.

    To tackle the first issue, our experiment not only has a larger sample size but also more meaningful data per
    sample. In addition, a deep learning model should be able to exploit this better in comparison to a multinomial
    naive Bayes classifier. The second issue is more fundamental. Since it is very likely that the \texttt{Server:} header
    labels follow a Zipfian distribution, the amount of target classes may run into the tens of thousands.
    In this work, we investigate some strategies to prepare a workable set of target classes.

    \subsubsection*{Can we use an NLP architecture to represent status lines as well as other HTTP header features?}
    The strategies we have outlined above represent several improvements over the experiment by \citet{book2013automated}.
    However, it still implies the use of a categorically encoded string of text
    that will, presumably, only have a couple of thousand unique combinations at most\footnote{As there are 63
    officially defined response codes, multiplied by 3 possible protocol versions, a ballpark calculation would
    indicate that one might expect 189 different combinations to occur in the data. However, one must also account
    for differences in formatting, such as whitespace use and casing.}.

    We propose using a state-of-the-art Transformer model to create dense, context-aware vector space embeddings to
    represent HTTP response headers. These embeddings preserve ordering information, which many of the research papers mentioned in section \ref{section2}
    show is a significant factor in the fingerprinting process.
    These embeddings can then be used in a neural network to perform multiclass classification.
    Although for this experiment we will only use the first line of the HTTP header,
    the technique should scale to entire headers. Using complete headers is unfortunately not possible in this work due to
    computational and storage constraints.

    \section{Background and related work} \label{section2}

    \subsection{Theory of web server fingerprinting}

    The reason why web server fingerprinting is possible in the first place, \citet{lee2002detecting} argue, is
    because of the nature of the documents that describe the protocol.
    They have some ambiguity in how they should be interpreted, and leave flexibility in how they should be implemented.

    \citeauthor{lee2002detecting} describe some of the characteristics that distinguish implementations from each other:

    \begin{itemize}
        \item Lexical: e.g., the response code message and capitalization, header wording and capitalization, and line
        terminators;
        \item Syntactic: e.g., the header ordering, ordering of lists in header values, and the formatting of values;
        \item Semantic: e.g., the presence or absence of specific headers, and the response given to malformed or undefined
        requests.
    \end{itemize}

    All that is required to craft a fingerprint for a web server is provoking it with a multitude of different
    requests, and recording its response to each. It should be noted that \citeauthor{lee2002detecting} argued that some of these
    characteristics could not be uniquely assigned to one of the three linguistic domains. In this work, we share
    their stance that, for the purpose of identifying characteristics for web server fingerprinting, we do not need to
    spend too much time debating which test case belongs to which domain.

    The topic of web server fingerprinting was revisited by \citet{yang2010improving}, who found that masking software\footnote{Such as ModSecurity and ServerMask.}
    was increasingly being used to protect against fingerprinting. As those tools are unable to completely mask all responses,
    the web server still leaks a certain set of characteristics, which become an indicator that masking software is used.
    They argue that the only comprehensive solution against fingerprinting is by making sure that
    there is no difference in how the HTTP specification is implemented by software.
    \citet{huang2015analyzing}
    described some additional fingerprinting characteristics

    and concluded that detecting individual software versions was
    challenging from HTTP status codes alone.
    \citet{book2013automated} provide the inspiration for this work.
    They crafted 10 different HTTP requests, and sent them to 120.000 domains in total.
    These were the highest-ranking 110.000 and the lowest ranking 10.000 sites from the
    Alexa Top 1 Million Sites dataset - the most popular websites on the
    Internet by descending order.

    As features, they used the returned HTTP status code for each test.
    The target label required for supervised machine learning was extracted from the \texttt{Server:} header, which
    was reported by the majority of servers.
    They found that server type prediction was possible, although their methodology was not powerful enough to distinguish individual versions.
    Their dataset was also highly unbalanced, as a single type constituted almost half of the data.
    For future work, they suggested that a more comprehensive suite of test cases should be designed, as they relied
    only on their informed judgement. They also recommended the use of more data
    as well as fingerprinting server software other than web servers.

    \subsection{Changes in the distribution of the World Wide Web}
    When the experiment by \citet{book2013automated} was performed, the use of transport-layer encryption was uncommon. It was
    only in 2016, after the introduction of Let's Encrypt\footnote{\url{https://letsencrypt.org/2016/04/12/leaving-beta-new-sponsors.html}}
    and the Automatic Certificate Management Environment (ACME) protocol by \citet{rfc8555},
    that TLS certificates became free of charge. In fact, there is no mention of HTTPS at all in the paper by \citet{book2013automated}. Concretely, this means
    that not only do we have to make requests to the default, unencrypted HTTP port 80, but also to HTTPS port 443 to
    get a complete picture of a typical, general-purpose web server.

    Additionally, a new version of the HTTP protocol was officially introduced by \citet{rfc7540}.
    HTTP/2 is limited to technical, low-level improvements, and does not introduce any changes in semantics.
    These low-level changes, however, may result in more diverse protocol noncompliance by server software, as the implementation and interpretation of the protocol is up to the developers.
    HTTP/1.1 is still typically used for unencrypted connections, and HTTP/2 is typically only used for encrypted connections.

    Modern web browsers generally negotiate HTTP/2 support in a backwards-compatible manner
    (i.e., servers are always given the opportunity to downgrade to a lower version of the protocol), and only attempt it
    over an encrypted connection. Web servers may thus vary in how well they support HTTP/2 under atypical conditions.
    These include not negotiating in a backwards-compatible way, or attempting to use HTTP/2 over an unencrypted connection.

    \subsection{Detection of malicious HTTP client requests}

    In the reverse context, which is to say detecting \emph{clients} from their \emph{requests} as opposed to
    \emph{servers} from their \emph{responses}, much more research has been done. This is an active and popular research topic,
    as spam and malware continues to evolve.
    Whereas server fingerprinting is mostly rule-based, machine learning models trained on HTTP data from client requests have been extremely successful
    in detecting malicious or anomalous clients, generally achieving between 92\% and 99\% accuracy
    \citep{yu2018attention, yuan2018detecting, niu2019using, laughter2020detection}.
    It should be noted that almost all of these experiments constitute a form of binary classification, which is generally easier
    than a multiclass problem such as detecting specific types or versions.

    A notable innovation was developed by \citet{mizuno2018detecting}. They built a system called BotDetector,
    to detect malware-infected clients from their HTTP
    traffic. Their central idea was to use a template generation algorithm based on DBSCAN clustering. Their
    algorithm clusters the keys and values of HTTP headers, resulting in fewer unique values, and thus fewer feature
    dimensions. The precision and recall of their models were high, indicating that this is a promising approach to
    reduce feature dimensionality. \citet{niu2019using} used a similar technique, but on a
    different dataset, and attempted to optimize XGBoost to perform classification. They found that XGBoost slightly
    outperformed the models tested by \citet{mizuno2018detecting}, and agreed that template generation is effective.
    Template generation is thus a potential method to reduce the large amount of target classes in our experiment.

    A number of experiments concentrated on creating (neural) embeddings and representations of HTTP headers
    instead of performing manual feature engineering \citep{li2020weighted, yuan2018url2vec, yu2018attention}. For example, by
    using tf-idf, word2vec, GLoVe, or combinations of those.
    The most recent of these experiments was performed by
    \citet{gniewkowski2021http2vec}. They used a RoBERTa model and BPE tokenizer to generate a representation model for
    HTTP headers, for the purpose of detecting anomalous requests.

    They trained their model only on unlabeled normal traffic,
    and found that simple classification algorithms could distinguish normal and anomalous traffic from the embeddings alone.

    Given the performance of embeddings, neural architectures, and machine learning in general
    in many of the studies in this field, the main takeaway for this work is that it seems to be feasible to
    treat HTTP header data as a natural language, with its own grammar and vocabulary\footnote{One might
    conclude that humanity's greatest technical infrastructure is fundamentally backed by the same thing as our social
    infrastructure: language.}. As almost all papers in the field highlight that ordering information, in particular when it concerns
    the representation of URLs and headers, is an important predictor, any strategy for HTTP representation would do well to model sequences as wel as occurrences.

    \subsection{Detection of malicious web servers}

    \citet{niakanlahiji2018phishmon} created a classical machine learning framework for detecting phishing web pages.
    They performed extensive feature engineering, and found that the presence or absence of certain HTTP header
    fields, the number of header fields, and the number of non-standard header fields were important indicators of whether a website was malicious or not.

    \citet{alhunting} developed two feature extraction methods to detect malicious web servers from HTTP
    response headers. They concluded that no good feature extraction methods existed, and that these are prerequisites
    for doing effective machine learning experiments. They noted that their research was closely related to web server
    fingerprinting. They sought to perform binary classification (malicious/non-malicious),
    and only relied on passive data (i.e., a single HTTP request)\footnote{As mentioned in the first section, software type
    and version fingerprinting typically requires actively provoking HTTP responses.}.
    They proposed a uniqueness value for each encountered header, and recording the indexed order in which headers
    occur for each sample.

    In their paper, they dismissed natural language processing techniques as a good way to tackle the problem. They
    argued that the sparsity and size of a tf-idf Bag-of-Words vector, along with the inability to capture ordering
    information, make it unsuitable.

    \subsection{Strategies on document representation}
    \citet{vaswani} introduced the Transformer model. It dispenses completely with the previous recurrent and convolutional approaches
    to sequence representation, and only uses multi-head attention. This architecture requires significantly less training time,
    as it is simpler and more parallelizable in nature.
    The Transformer model is extremely popular because of this, and has a wide range of applications including, and not limited to,
    machine translation, question-answering systems, classification, and named entity recognition.
    \citet{gniewkowski2021http2vec} have demonstrated their viability in representing sequence-aware HTTP headers, and dispenses with the need
    to use a bag-of-words approach, which \citet{alhunting} found unsuitable.

    Most modern of-the-shelf NLP frameworks are centered around the idea that a single sample is a single text document,
    but this is not the case for our experiment. Our samples are lists of individual documents, of which each document is
    - in principle - a complete HTTP response header with a typical length of \textasciitilde100-1500 characters of plain text.
    The maximum token length supported by popular Transformer
    models such as BERT is 512, meaning that longer texts must be split up or truncated. Ideally, we would like to
    represent our samples as one continuous document split by control tokens, but that is computationally infeasible as
    it would result in over 10k tokens.
    This is not possible even with long-sequence Transformer models such as the Longformer \citep{beltagy2020longformer} and Big Bird
    \citep{bigbird2020}.
    Furthermore, we do not necessarily need to retain sequence information between test cases in the first place, as the
    HTTP requests are more or less independent from each other. It is likely sufficient to encode each test case
    separately, and then concatenate the resulting vectors to represent a single sample.

    Transformer models are usually fine-tuned on a downstream task such as classification by attaching a different
    head to the last hidden layer.
    Given that we must encode our test cases separately, this approach will not work.
    Instead, we use the final hidden state of the Transformer as input for a separate feed forward neural network.
    \citet{reimers-2019-sentence-bert} note, however, that the raw, pooled final hidden state may not be a good
    representation of a document as a whole. In ordinary Transformer classifiers, this is not a
    problem as the encoder is immediately fine-tuned on a downstream task.

    When using the output of a Transformer encoder directly as an embedding,
    \citet{reimers-2019-sentence-bert} suggest fine-tuning it on a semantic textual similarity task to generate superior
    semantic embeddings. This unfortunately requires an annotated corpus. There have been recent advances
    in unsupervised sentence embedding learning as well, for example by \citet{wang-2021-TSDAE}.

	    \section{Data and Material} \label{section3}
    In order to perform any sort of machine learning experiment and answer our research questions, we require a dataset.
    As no (public) web server fingerprinting dataset exists, we constructed our own. In this section, we
    describe how we gathered the data and pre-processed it, along with details on our fingerprinting test suite.

    \subsection{Test cases}
    The test cases we use to construct fingerprints from web servers have been designed based on strategies discussed
    in the literature by \citet{lee2002detecting}, \citet{book2013automated}, \citet{huang2015analyzing}, \citet{yang2010improving},
    and those employed by existing rule-based fingerprinting tools. We tried to find a good balance between
    exhaustively trying all
    thinkable interesting scenarios, and minimizing bandwidth, storage, and model complexity. We list the specifics
    of each test in appendix \ref{sec:testcases}.

    In almost all circumstances, the \texttt{Host:} and
    \texttt{Accept:} headers are expected as part of a valid HTTP request. Many web servers host multiple websites,
    and the \texttt{Host:} header is used to disambiguate which specific website is requested. The \texttt{Accept:}
    header indicates what type of content is acceptable for the client. These two headers, along with
    \texttt{User-Agent:}, are always sent with standard HTTP-request tools such as \texttt{curl} and \texttt{wget}, and we
    therefore believe it is reasonable to include those headers by default. The HEAD method is identical to
    a GET request, but instructs that the server should not actually send any actual website content. The HTTP/1.1 and HTTP/2 specifications
    indicate that the HEAD method \emph{must} be supported by compliant web servers. Using HEAD instead of GET is bandwidth-efficient,
    and we use it as the default request method for that reason.

    We only perform requests that are marked as safe and idempotent by the HTTP specification, and ones
    that will plausibly not leak private information.
    Concretely, this means that we will limit ourselves to the GET, HEAD, and OPTIONS methods, as
    \citet{book2013automated} did. We will also not attempt a path traversal attack by making relative requests,
    even though this was suggested as a fingerprinting strategy by \citet{lee2002detecting} and implemented by
    \citet{book2013automated}, as it may leak private information. In Figure \ref{fig:requestexample}
    an example of an HTTP HEAD request can be seen. This example is equivalent to the baseline (test case 0)
    directed at the \texttt{www.rug.nl} domain.

                \begin{figure*}
                            \caption{Example of a raw, plain-text HTTP HEAD request.\label{fig:requestexample}}
        \begin{verbatim}
HEAD / HTTP/1.1\r\n
Host: www.rug.nl\r\n
User-Agent: Mozilla/5.0 (X11; Linux x86\_64; rv:105.0) Gecko/20100101 Firefox/105.0\r\n
Accept: */*\r\n
\r\n
    \end{verbatim}
    \end{figure*}

    Notably missing from the test suite are (malformed) absolute URLs in the first line of the
    request~\citep{lee2002detecting}, and requests of a large size~\citep{lee2002detecting, huang2015analyzing}.
    We chose not to include these as we would have to test multiple absolute URLs to get meaningful differences for the first scenario.
    For the second scenario,  we would have to try at least four tests with arbitrarily chosen URL lengths according to        \citet{lee2002detecting} and \citet{huang2015analyzing}.
    We therefore consider these two scenarios to be too time and storage expensive.

    \subsubsection{Testing HTTP/2 support}
    \nocite{10.17487/RFC1945}
    We perform four requests which specifically evaluate HTTP/2 support. These tests are on two axes: whether it supports
    the protocol encrypted and/or unencrypted, and how the server behaves when support is assumed
    (i.e., when HTTP/2-specific packets are immediately sent) versus when it is negotiated (i.e., the server is given the opportunity
    to "downgrade" to a lower version of the protocol).

    \subsection{Collector algorithm}
    In this section, we describe the algorithm that we devised in order to gather our raw data. As input to our algorithm,
    we use domains from the Tranco list by \citet{pochat2018tranco}. This is a robust ranked meta-list of millions of domains,
    not unlike the Alexa dataset which \citet{book2013automated} used.
    We were able to gather data for the top 4.8 million domains in that list, over the span of two days.

    \subsubsection{Overall architecture}
    All operations time out after 2 seconds. If we do not get a valid reply from a server within that time, we
    assume that it is down and note an \texttt{<ERROR>} token for that specific test. We also do not verify TLS
    certificates, as having an actually secure connection does not matter for our experiment.

    We submit each domain in the input list to a thread worker, and thus perform parallel processing. To maximize
    performance, all thread workers are completely asynchronous. When a worker has finished, it will
    immediately append the tab-seperated output of the domain to a file. It will then proceed with the next domain in
    the queue.

    \subsubsection{Resolving and locating websites}
    First, a domain is resolved to an IPv4-address. If this fails, we assume that the domain is dead and assign
    \texttt{null} values to all the tests.
    If resolving the domain succeeds, we attempt to establish a canonical location for the website\footnote{The canonical location
    of a website is the location that we end up with after following location redirects.}.
    We follow redirects (i.e., \texttt{Location:} headers) up to a maximum of three hops. Note
    that we only follow redirects to the same port and protocol. For example, \texttt{http://google.com} will be
    followed to \texttt{http://www.google.com}, but will not be followed to \texttt{https://www.google.com}.

    Some servers only listen on either HTTP/80 or HTTPS/443. If we fail to establish a canonical location for one of
    the ports, we assign \texttt{null} values to all the tests of that port.
    If we can establish a canonical location, we use the output of that procedure as test case 0.

    \subsubsection{Running the tests}
    After establishing the canonical location for a domain, we run all tests on each port, if that port is reachable.
    We send our HTTP/0.9, HTTP/1.0, and HTTP/1.1 requests as a byte-encoded string through a raw socket connection,
    and read the first 4096 bytes of the TCP response. Note that we wrap this raw socket in an encryption layer for
    the HTTPS/443 tests.
    It is non-trivial to implement the new
    state-of-the-art HTTP/2 protocol manually. Whereas the older versions are purely textual, HTTP/2 is in a binary format and requires compression of the headers.
    We therefore use the \texttt{curl} command line tool to perform HTTP/2 requests.
    In total, our test suite results in 32 HTTP requests: 16 for each port (80 and 443).

    \subsubsection{Processing HTTP responses}
    Internally, the HTTP response is stored both as a raw string, and as a case-insensitive dictionary of headers.
    From this, we extract the status line (e.g.,\texttt{HTTP/1.1 200 OK}) and \texttt{Server:} header (e.g.,\texttt{nginx/1.18.0}).
    If the first header (i.e., the status line) is longer than 100 characters, we assume that we are dealing with an
    HTTP/0.9 compliant HTML-only response and substitute the content by an \texttt{<html>} token.
    This is because the status line generally does not exceed 75 characters.
    If we encounter an error while parsing the response, we assume that we are dealing with a malformed or
    non-compliant response, and we substitute the status line by an \texttt{<ERROR>} token.
    If there is no \texttt{Server:} header, we substitute it by an \texttt{<EMPTY>} token.
    In Figure \ref{fig:responseexample} an example of a HTTP response can be seen.

                \begin{figure*}
                            \caption{Example of the raw, plain-text HTTP response to the HEAD request from Figure \ref{fig:requestexample}.
                            The original request was sent over a port 80 TCP socket connection.\label{fig:responseexample}}
        \begin{verbatim}
HTTP/1.0 302 Moved Temporarily\r\n
Location: https://www.rug.nl/\r\n
Server: BigIP\r\n
Connection: Keep-Alive\r\n
Content-Length: 0\r\n
\r\n
    \end{verbatim}
    \end{figure*}

    \subsection{Annotation}
    We use two types of target labels: major and minor. The major label is the web server type, such as \texttt{nginx},
    and the minor label is the web server type and version combined, such as \texttt{nginx/1.18.0}.
    As mentioned in the introduction, selecting and creating such target labels is non-trivial. Some of the
    issues are:

    \begin{itemize}
        \item the number of different web servers in existence;
        \item the differences in formatting (e.g., capitalization; use of special characters; spacing);
        \item the Zipfian distribution of web servers (i.e., the dataset is highly unbalanced, with the tail consisting of
        web servers with only a handful of examples).
    \end{itemize}

    Below, we discuss some of the strategies to tackle these issues.

    \subsubsection{Class combination}
    To combine classes, one may apply naive preproccesing such as lowercasing and removing special characters.
    This removes some of the differences in formatting. Alternatively, one might use a clustering algorithm
    to group very similar classes based on frequency, similar to what \citet{mizuno2018detecting} and \citet{niu2019using} did.
    Their template generation algorithm is promising, but unfortunately non-trivial to implement.

    \subsubsection{Class reduction}
    \citet{book2013automated} only inspected the versions of a single software type, \texttt{Apache}. This could be extended
    to the top-\emph{n} major software types. Another method would be to use \emph{almost} all labels,
    but exclude samples with a frequency lower than two or three standard deviations.
    One might also use all samples and corresponding labels, and see if this significantly affects training time and evaluation performance.

    \subsection{Data distribution}

    In Table~\ref{tab:web-server-distribution} we make a comparison between the distribution of server types, as reported by different organisations.
    The distribution of our target labels appears to be similar to \citet{book2013automated}, and general surveys
    done by Netcraft\footnote{\url{https://news.netcraft.com/archives/2022/11/10/november-2022-web-server-survey
.html}} and w3techs\footnote{\url{https://w3techs.com/technologies/overview/web_server}}. As the methodologies of
    the sources vary, direct comparisons cannot be made and the table should be considered to contain a rough approximation only.

\begin{table*}
        \caption{Distribution of the top-5 general-purpose web servers, excluding known proxies, CDNs, caches, and
        proprietary, non-public, organization-specific software.}
        \label{tab:web-server-distribution}
        \resizebox{\textwidth}{!}{
            \begin{tabular}{@{}lllllll@{}}
                \toprule
                \textbf{Web server} & \textbf{Us} & \textbf{\citeauthor{book2013automated} top 100k} &
                \textbf{w3techs} & \textbf{Netcraft all} & \textbf{Netcraft active} & \textbf{Netcraft busiest} \\
                \midrule
                \textbf{Apache} & \textbf{20.7\%} & 40.7\% & 31.5\%
                & 21.4\% & 21.6\% & 21.7\% \\
                \textbf{nginx} & \textbf{20.7\%} & 12.8\% & 34.2\%
                & 26.5\% & 19.6\% & 21.2\% \\
                \textbf{LiteSpeed} & \textbf{4.4\%} & 1.5\% & 12.3\%
                & 5\% & 5\% & 3\% \\
                \textbf{Microsoft-IIS} & \textbf{2.9\%} & 12.7\% & 5.9\%
                & 3\% & 3\% & 5.4\% \\
                \textbf{openresty} & \textbf{2.1\%} & n/a & n/a
                & 8.1\% & 3\% & 1\% \\ \bottomrule
            \end{tabular}
        }
    \end{table*}

    \subsection{Processing}
    For this experiment, we have chosen an uncomplicated approach to sample and label filtering. The five major server types we use in this
    experiment are popular, publicly available, general purpose origin servers which make up approximately between 40\% and 60\% of the World Wide Web. As caches, content delivery
    networks, load balancers, and proxies may unpredictably affect the result, we have chosen to omit those entirely\footnote{This is based on self-advertised \texttt{Server:} headers. It is possible some samples are behind one of those systems but do not advertise it.}.
    This causes our dataset to be unrepresentative of the true distribution of the World Wide Web, as a part of our raw data is removed.
    For example, the Cloudflare CDN constitutes a sizable 13\% of the data.
    After removing poor samples and filtering the target classes, approximately 1.96 million out of 4.77 million
    samples remain. 24\% of these contain a minor label. The filtering of classes and samples are explained below.

    \subsubsection{Filtering target classes}
    We apply two simple regular expressions to filter the samples. Note that everything is lowercased before filtering.

    A sample is dropped if it does not match \texttt{(\textasciicircum vendor)}, where vendor is any one of
    \texttt{apache}, \texttt{nginx}, \texttt{litespeed}, \texttt{microsoft-iis}, or \texttt{openresty}. If a label
    matches, an additional filter is applied: \texttt{(vendor/[0-9\textbackslash.]+)}. If it matches this additional filter,
    the matching part is used as the version label. Note that this method accounts for labels such as
    \texttt{nginx/1.18.0 (Ubuntu)} and \texttt{Apache/2.2.24 (Unix) mod\_ssl/2.2.24 OpenSSL/1.0.1e-fips PHP/7.2.23
    mod\_pubcookie/3.3.4a mo\_uwa/3.2.1}\footnote{Long labels like these frequently occur in the data.}. This results
    in 347 unique vendor/version combinations\footnote{LiteSpeed never reveals minor versions, and is thus excluded
    from the minor version experiment.}.

    If the first expression is matched, but the second is not, the sample is given the \texttt{<MAJOR>} token to
    indicate that it does not reveal a minor version.
    The dataset can be easily adapted to different major and minor classes by re-running the pre-processing script
    and modifying the regex filters.

    \subsubsection{Filtering poor samples}
    We only keep samples that match the following criteria:

    \begin{itemize}
        \item contains a maximum of 50\% \texttt{<ERROR>} tags and empty strings in their responses;
        \item the most frequently returned \texttt{Server:} header is returned in at least 50\% of the responses, and is not an empty
        string or \texttt{<ERROR>} tag;
        \item it has the same most frequent \texttt{Server:} header for both HTTP/80 and HTTPS/443.
    \end{itemize}

    Our reasoning is that some structural errors or empty responses may in fact be fingerprints. If it happens too
    often, however, it is reasonable to assume that there is a malfunction or poor connection.
    Furthermore, if the returned \texttt{Server:} header is not identical for both ports, it is plausible that
    we are dealing with an attempt to evade fingerprinting, or that the requests are actually handled by different servers.

    \subsubsection{Generating sentences for tokenization and masked language modelling}\label{subsec:embeddinglist}

    As many responses are identical, removing duplicates is efficient for training and storage purposes. We take all
    unique responses to the test cases from the complete, non-filtered dataset, and remove some of the pure HTML lines (e.g., that start with <html><) as those are overrepresented.
    This results in a list of 3828 unique, real-world HTTP status lines and simple HTML responses. Interestingly, many of the status lines are quite
    different from what is recommended by the specification. Some of them are not in English, vary in capitalization,
    or are wholly custom. Some of the interesting ones are listed in appendix \ref{sec:interestingstatuslines}.

    It is also apparent that some HTTP/0.9-compliant HTML-only replies reveal the server type, because they serve a
    default page such as \texttt{<html>Apache is functioning normally</html>}. We considered explicitly replacing
    these entries, but given the relative rarity of these replies we reasoned it would not affect our results too
    much\footnote{In the results section, we show that feature importance as determined by a Random Forest confirms
    that this is the case.}. Furthermore, it could be argued that this represents a security leak, and not an
    oversight in data cleanup.

    \section{Method} \label{section4}

    Our overall strategy to answer the research questions is the following: we train a tokenizer and Transformer
    encoder from scratch on the list of status lines mentioned in section \ref{subsec:embeddinglist},
    use them to create dimensionality-reduced embeddings for all test
    cases of a sample, and concatenate those embeddings such that the entire sample is represented by a single vector.
    A feed forward neural network and a Random Forest classifier are then trained and evaluated on the randomized
    dataset to predict major types. The FNN is additionally trained and evaluated on minor version prediction.

    \subsection{Transformer model and tokenizer}
    We used the RoBERTa model by \citet{https://doi.org/10.48550/arxiv.1907.11692} as implemented by
    \citet{DBLP:journals/corr/abs-1910-03771}, and its corresponding byte-level BPE tokenizer. All settings were left
    at their defaults\footnote{See the \texttt{transformers 4.23.1} package and \texttt{roberta-base} configuration
    .}, except for the vocabulary size, which was set to 2048 instead of 30522. This seemed appropriate as the vocabulary of HTTP
    headers is significantly smaller than human language. RoBERTa was chosen as it achieves greater performance than BERT in most benchmarks,
    and was used by \citet{gniewkowski2021http2vec}.

    The model and tokenizer were trained from scratch on the unique sentence list described in the previous section.
    The training objective for the model was masked language modelling, with a masking rate of 0.15. MLM was introduced
    by \citet{devlin-etal-2019-bert} and is an unsupervised auto-encoding method to pre-train Transformer architectures
    with context-aware linguistic information, after which the model can be used in a downstream task.
    Individual tokens in the input documents are randomly masked, and the model is then trained to fill in the missing token.
    Our model did this for 30 epochs, after which it had a loss of 1.17.
    The validation plot indicated that we had reached a point of diminishing returns.
    Considering the simplicity of the language (HTTP status lines only) and the risk of overfitting,
    we found further training not to be necessary for this particular experiment.

    \subsection{Embedding generation}
    Having trained the Transformer encoder, we used it to generate a vector-space representation for a single document (i.e., status line from a single test case).
    This document is tokenized with padding and truncation, and fed through the RoBERTa model.
    For this we disable gradient calculation and apply mean pooling on the final hidden layer as
    implemented, and recommended by, the SentenceTransformers project\footnote{\url{https://www.sbert.net/examples/applications/computing-embeddings/README.html\#sentence-embeddings-with-transformers}}
    \citep{reimers-2019-sentence-bert}. This results in a raw, 768-dimensional sentence-level embedding.

    We do this for all lines in the unique sentence list (section \ref{subsec:embeddinglist}) and fit a 64-component principal component analysis model on them.
    We chose to implement dimensionality reduction through PCA, as it was suggested by the
    SentenceTransformers project\footnote{\url{https://www.sbert.net/examples/training/distillation/README.html\#dimensionality-reduction}}. They found that this has a very limited impact on
    subsequent model performance on semantic textual similarity tasks, while significantly reducing computational and storage complexity.

    Then, for each sample in the dataset:
    \begin{itemize}
        \item we generate pooled embeddings for all 32 test case responses individually;
        \item we apply a PCA transform on those 768-dimensional vectors to reduce them to 64 dimensions;
        \item we concatenate those embeddings into a single 2048-dimensional vector.
    \end{itemize}

    A ready-to-use dataset is then created. Each sample contains the major class label, minor class label, and a
    2048-dimensional float32 array.
    This dataset is randomized, and split into a train set (80\%), validation set (10\%), and test set (10\%).

    \subsection{Classifiers}

    \subsubsection{Feed forward neural network}

    There are no formal strategies for designing a feed forward artificial neural network. According to
    \citet{GULIYEV2018262}, two hidden layers are required to model functions of any shape. The ReLU activation
    function \citep{https://doi.org/10.48550/arxiv.1803.08375}, Adam optimizer \citep{https://doi.org/10.48550/arxiv.1412.6980}, and use of dropout layers \citep{JMLR:v15:srivastava14a} after each hidden layer are considered good practice, and halving
    the amount of dimensions after each layer is a good rule of thumb \citep{10.5555/1502373}. The batch size was 1000, and the
    loss function was sparse categorical crossentropy. Conceptually, the model is a multilayer perceptron. It is implemented through the
    Keras API, with the following layers:

    \begin{enumerate}
        \item Input layer (2048 dimensions)
        \item Fully connected layer (1024 dimensions, ReLU activation)
        \item Dropout layer (rate 0.2)
        \item Fully connected layer (512 dimensions, ReLU activation)
        \item Dropout layer (rate 0.2)
        \item Output layer (5 or 347 dimensions depending on classes, softmax activation)
    \end{enumerate}

    Having carefully monitored the validation loss, it was determined that the point where there was no significant
    gain in performance was at 20 and 30 epochs, for major type prediction and minor version prediction respectively.

    \subsubsection{Random Forest}
    We used a Random Forest classifier (i.e., ensemble of decision trees) as implemented by the \texttt{scikit-learn 1.1.3} library with the default settings. It
    was trained and tested on a subset of the data (5\% of the train set; 50\% of the test set) as the computational complexity of the model is large.
    An ensemble of decision trees can learn non-linear relationships and provides interpretable, impurity-based Gini importance for
    all features.
    This enables us to investigate which test cases are meaningful, and serves as a check that our
    classifiers learn from the test cases and that the data is not flawed or tainted.
    Furthermore, it allows us to evaluate the benefits of a neural classifier compared to a classical machine learning algorithm.

    \section{Results and Discussion} \label{section5}
    In this section, we first present a visual representation of the encoded web servers.
    We then report the classification metrics and the baseline f1-scores for both classification tasks, and compare them with \citet{book2013automated} as well as we can.
    In addition, we evaluate the discriminative power of our test cases.
    As there is very little prior public research on web server classification, we have no benchmarks to directly compare against.

    \subsection{Transformer embeddings}

\begin{figure*}
				\centering
        \includegraphics[width=6in]{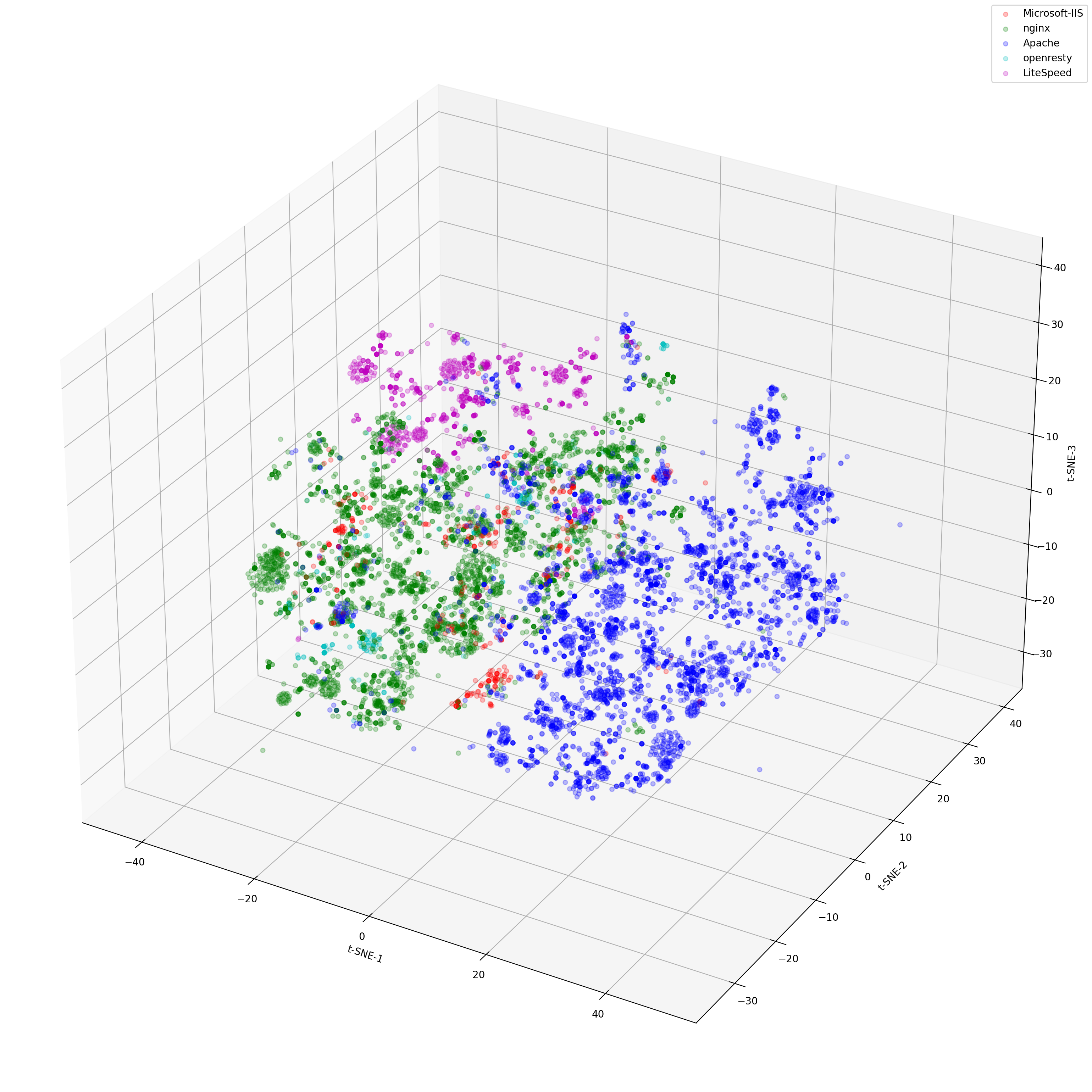}
        \caption{3-dimensional T-distributed Stochastic Neighbor Embedding \citep{7b54165e73a3424b8820136bcf61ca89} of 10.000 random samples, colored by major server type.\label{figure:tsne}}
    \end{figure*}

    See Figure~\ref{figure:tsne} for a visualization of the dataset. These are the Transformer-encoded, PCA-reduced, concatenated representations of individual web servers. There are clearly visible clusters discernible to the human eye, indicating that the test case response data and generated embeddings are appropriate for distinguishing between web servers.

    \subsection{Server type classification}
    In Tables \ref{tab:classification-report-ffnn} and \ref{tab:classification-report-rf} we show the classification reports for both classifiers. In these tables, support is the amount of samples in the test set which have that class as the true label. In appendix \ref{sec:cms} we list Figures \ref{figure:ffnn_majortypes_testset_abs}, \ref{figure:ffnn_majortypes_testset_normalized}, \ref{figure:rf_majortypes_testset_abs}, and \ref{figure:rf_majortypes_testset_normalized}, which are the confusion matrices for the respective classifiers.

    We establish our baseline scores by evaluating a dummy classifier, which infers the most frequent class
    from the train set and then always predicts that class on the test set. This represents a naive but clever guessing strategy.
    Any classifier must be able to beat this baseline for it to be considered "better than random".

\begin{table*}
\centering
\begin{tabular}{@{}lllllr@{}}
    \toprule
                      & precision & recall & \textbf{f1-score} & & support   \\ \midrule
Microsoft-IIS         & 0.95      & 0.91   & 0.93              & & 7911      \\
nginx                 & 0.96      & 0.98   & 0.97              & & 80571     \\
Apache                & 0.98      & 0.98   & 0.98              & & 83185     \\
openresty             & 0.96      & 0.85   & 0.90              & & 5655      \\
LiteSpeed             & 1.00      & 1.00   & 1.00              & & 18481     \\ \midrule
\textbf{accuracy}     &           &        & \textbf{0.98}     & & 195803    \\
\textbf{macro avg}    & 0.97      & 0.94   & \textbf{0.96}     & & 195803    \\
\textbf{weighted avg} & 0.98      & 0.98   & \textbf{0.98}     & & 195803    \\ \bottomrule
\end{tabular}
\caption{Classification report of the FNN classifier.}
\label{tab:classification-report-ffnn}
\end{table*}

\begin{table*}
\centering
\begin{tabular}{@{}lllllr@{}}
    \toprule
                      & precision & recall & \textbf{f1-score} & & support \\ \midrule
Microsoft-IIS         & 0.94      & 0.88   & 0.91              & & 3900    \\
nginx                 & 0.96      & 0.97   & 0.97              & & 40307   \\
Apache                & 0.98      & 0.98   & 0.98              & & 41563   \\
openresty             & 0.94      & 0.82   & 0.87              & & 2877    \\
LiteSpeed             & 1.00      & 1.00   & 1.00              & & 9255    \\ \midrule
\textbf{accuracy}     &           &        & \textbf{0.97}     & & 97902   \\
\textbf{macro avg}    & 0.96      & 0.93   & \textbf{0.94}     & & 97902   \\
\textbf{weighted avg} & 0.97      & 0.97   & \textbf{0.97}     & & 97902   \\ \bottomrule
\end{tabular}
\caption{Classification report of the Random Forest classifier.}
\label{tab:classification-report-rf}
\end{table*}

    We consider macro f1-score to be a good evaluation metric, as it compensates for the class imbalance and dominance of \texttt{nginx} and \texttt{Apache}. The baseline macro f1-score for this task is 0.12. \citet{book2013automated} did not report any of the common classification metrics, and tested their Bayesian model against a portion of their training data,
    instead of a reserved test set. Although not explicitly described as such\footnote{They described their accuracy somewhat confusingly as: "\emph{our algorithm produced different results from the server type returned by the site for approximately 38\% of sites}". The most frequent type occurred in 41\% of the cases, which makes their result better than majority guessing, but not enormously so.}, it appears that they achieved 62\% accuracy with 8 classes.

    Overall, both the FNN and RF classifier perform beyond all expectations, achieving 0.96 and 0.94 macro f1-score respectively. This suggests that all five web servers have a distinct set of fingerprints, which are apparent from status lines alone. As a simple RF classifier is able to discern this signal with remarkable accuracy, it is likely that these are surface-level fingerprints and are not necessarily learned from subtle, weakly defined relationships between test cases.

    Both classifiers struggle with detecting \texttt{Microsoft-IIS} and \texttt{openresty}, relatively speaking. This matches with what is seen in Figure~\ref{figure:tsne}: these two classes are less well-defined than Apache, nginx, and LiteSpeed. It is, however, unknown whether this is caused by the limited sample size of those classes or because they are inherently more ambiguous.

    Furthermore, it is possible that our performance is the ceiling of what is achievable given the possibility of a reported target label being forged. Forging may have been done manually, or through masking software. If a target label is forged, it cannot be expected that a system can reliably classify it, given that it is trained on the true signal that belongs to that class.

    \subsection{Server version classification}
    The minor version classification task was only performed by the FNN. The corresponding classification report can be seen in Table \ref{tab:classification-report-ffnn-minortypes} in appendix \ref{sec:classificationreports}. Note that due to the rarity of some target classes, they may not appear in all three splits, and may not have been learned by the classifier. Additionally, very rare classes may be forged\footnote{For example, \texttt{microsoft-iis/99.99} has not been released by Microsoft but is present in the list of classes.}. For these reasons, we consider the weighted f1-score to be a better metric than the macro f1-score for this task. The macro f1-score compensates for class imbalance by taking the unweighted mean of each label's f1-score, and thus treats each class equally. The weighted f1-score, on the other hand, weighs the f1-scores by the number of true instances for each label, and thus assigns a greater importance to more frequent labels.
    In this case, more frequently occurring labels are more likely to be authentic.

    The baseline weighted f1-score on this challenging 347-class classification task is 0.02. Given that only the first line of the HTTP header is used in each test case, a weighted f1-score of 0.55 for a FNN is impressive. This is a non-trivial improvement in classification performance.

    Our hypothesis was that it would not be possible to discriminate minor versions from status lines alone.
    Clearly, this is wrong.
    It is plausible that, given complete headers, more data, and longer training time for the Transformer encoder and feed forward classifier, better performance can be achieved.
    There is no apparent reason why our proposed method would not scale to more classes and more data.

    \subsection{Test case and feature analysis}
    In Figure \ref{tab:feature-importances} in appendix \ref{sec:featureimportances}, the calculated Gini impurity can be seen. The Random Forest classifier provides this metric on a per-feature basis. We divided the 2048 features by 64, rounded them down to the nearest integer, and took the mean of the resulting values to construct an interpretable average per test case.

    Tests run on HTTPS/443 appear to be more important than those run on HTTP/80 overall, although the relative ranking of tests appear to roughly align. This matches with what is seen in the data: many HTTP/80 responses are 3xx location redirects that point towards HTTPS/443. This is also reflected in the discrepancy in importance between protocols for test cases 12-15. These tests depend on whether actual content is served: location redirects do not have those.

    The most powerful predictor is test case 7. Looking at the data, \texttt{Apache} almost always causes an \texttt{<ERROR>} token, whereas the other web servers generally return \texttt{404 Bad Request}. The second most powerful predictor, test case 2, appears to fingerprint \texttt{Microsoft-IIS}. This web server never supports HTTP/2 on port 80: this target class is thus ruled out if any response at all is given.
    Test cases 8 (invalid HTTP version) and 9 (invalid method) were suggested as fingerprints by almost all previous papers. The high Gini impurity of these features thus matches the literature.
    Explicit support for HTTP/1.0 as tested by cases 4 and 5 is also an important predictor.
    Furthermore, test case 0 is the lowest ranked test for both protocols. This matches what is expected from a baseline, standard HTTP request.
    Test case 15 appears to hold no predictive power on both protocols.
    It is possible that it is too dependent on the actual content of a website to be a meaningful discriminator of web servers.

        Given that the learned feature importances match our intuitions and the general patterns in the data, it is likely that our data collection, processing, and embedding were done correctly. There are no odd or otherwise unexpected results.

    \section{Conclusion} \label{section6}
    The goal of our study was to investigate whether state-of-the-art deep learning, big data, and natural language processing techniques
    could be used to improve the ability to detect vulnerable web server versions.
    To answer our first research question: by performing classification on Transformer-encoded HTTP response headers, we have demonstrated that it is possible to achieve remarkable precision and recall beyond what was achieved by \citet{book2013automated} and by naive guessing. This is a clear indication that machine learning and big data approaches are promising strategies for performing software fingerprinting when static code analysis is not available, and only black-box input/output data can be used. This is a significant discovery in the context of software vulnerability research.
    It is currently unknown which specific improvement over the original experiment is chiefly responsible for the improved classification performance. That said, the combination of more samples, more test cases, testing multiple protocols, and more advanced machine learning algorithms is a successful one.

    Our second research question asked whether a classifier could discriminate specific server versions. We have been able to detect which type and version of server software is used with unprecedented precision and recall, based on status lines alone. Although the fundamental issue of Zipfian class sparsity remains unsolved, our naive pre-processing strategy results in a workable amount of target classes. These target classes make up approximately half\footnote{As mentioned earlier, sources disagree on the exact distribution of the World Wide Web.} of the web servers, which is a non-trivial amount.

    Lastly, our third research question asked whether we could use an NLP architecture to represent status lines as well as other HTTP header features, in the context of fingerprinting web servers. Based on the obtained f1-scores, our proposed tokenizer, Transformer encoder, and dimensionality reduction and concatenation strategy appears to result in an excellent representation of a web server. Although we were not able to test the encoding of complete HTTP response headers in practice, our system is designed in such a way that it should scale without any fundamental issues.

        \subsection{Limitations}

        Although the five selected target classes make up roughly half of the World Wide Web, they are only a small subset of all web server types in existence. It remains unknown whether precision and recall of rare server types will be similar. The proposed system should scale to many classes and work with entire headers, but this is yet to be demonstrated in practice.

    It is important to note that there are some fundamental issues with our automatic data collection.
    These were also identified by \citet{book2013automated}, except for a population sampling issue they did not
    consider. We have also not performed a thorough statistical analysis of our data, and thus cannot draw conclusions about the
    fundamental correctness and integrity of our dataset.
    Overcoming the limitations of the data is possible, but requires careful selection, annotation and verification of its
    correctness. The amount of manual work required for that is significant, and makes us wonder
    whether it doesn't defeat the original purpose of a machine learning approach requiring less effort than a rule-based system.

    \subsubsection{Different servers for different requests}
    In principle, it is entirely possible that different software listens on each port, or worse: each request may be
    handled by different software and/or hardware.
    There is, in fact, no guarantee at all as to how independent requests are handled by the underlying infrastructure.
    It would be reasonable to assume that only the highest-ranking domains would have the complex technical
    infrastructure required for multi-server and multi-software setups, which form only a small portion of the data.
    We leave determining a better method to detect this to future work.

    \paragraph{Mitigation:} Checking whether the returned \texttt{Server:} label is consistent and identical for both
    HTTP/80 and HTTPS/443.

    \subsubsection{The same server for different domains}
    It is also possible that a single server hosts the websites of multiple domains.
    This may cause a single web server to appear multiple times in the data, with potentially identical responses.

    \paragraph{Mitigation:} None.
    Although it is possible to filter by IP-addresses, we are unsure whether this limitation constitutes a natural occurring feature of the World Wide Web or a sampling problem in the data.

    \subsubsection{Content Delivery Networks, load balancers, caches, and proxies}
    Many websites are proxied for performance and anti-spam reasons. These may manipulate the \texttt{Server:} header
    as \citet{book2013automated} found. In principle, these relays should be transparent, in the sense that they are not
    supposed to manipulate the semantics such as status lines and the headers of replies.

    For this experiment, we have left them out. They are, however, a fundamental part of the World Wide Web as of writing.
    Further research should include them and report on their effects on fingerprinting performance.

    \paragraph{Mitigation:} Explicitly filtering our target labels to only include known origin servers.

    \subsubsection{Incorrect \texttt{Server:} headers}
    There is no guarantee that the returned \texttt{Server:} headers are correct. They may not only be simply
    left-out, but intentionally wrong as well. Unless we manually contact the server administrators, there is no
    guarantee that our target labels are trustworthy at all.

     \paragraph{Mitigation:} None. Like \citet{book2013automated}, we have assumed that these form a non-significant part of the data.

    \subsubsection{Dynamic \texttt{Server:} headers}
    \citet{book2013automated} found that some servers returned varying \texttt{Server:} headers for each request.
    Some appeared to be intentional to mislead attackers, and some were unintentional and caused by misconfigurations.
    Critically for web server fingerprinting, some misconfigurations were an attempt to hide or manipulate the \texttt{Server:}
    header but were not applied to all test cases and therefore constitute an exploitable leak.
    We leave determining how this can be exploited by a fingerprinting system to future work.

    \paragraph{Mitigation:} Checking whether the returned \texttt{Server:} label is consistent and identical for both
    HTTP/80 and HTTPS/443.

    \subsubsection{Masking software}
    As mentioned in the background section,  \citet{yang2010improving} found that masking software such as
    ModSecurity and ServerMask are increasingly used to protect against fingerprinting. However, they found that
    such software created a unique fingerprint by itself. We leave analyzing the effects and signatures of masking software to future work.

    \paragraph{Mitigation:} None. We have assumed that these form a non-significant part of the data.

    \subsubsection{Mismatch between the population that we want to detect and the population which we train on}
    This was not considered by \citet{book2013automated}, but it constitutes a fundamental problem with this
    methodology.
    We can, by the nature of this method of data collection, only infer the target labels from the sample of server
    administrators who do not bother to remove their \texttt{Server:} header.
    Of course, this sample is not necessarily interesting to fingerprint in the first place, as they freely advertise
    their \texttt{Server:} header anyway\footnote{Regarding the possibility of incorrect labels,
        \citet{book2013automated} argued that it was still interesting to calculate the confidence that a prediction
        matches the supposed true label, i.e., the probability that a header is falsified assuming that the classifier
        is reasonably accurate.}.
    This raises the critical question whether these populations actually come from the same distribution. It is
    certainly possible that administrators who remove \texttt{Server:} headers also employ additional
    anti-fingerprinting techniques, or use different (or newer) server software infrastructure altogether.
    Future research could focus on statistical analysis of these groups, to determine if there are fundamental differences.
    And if so, the implications of that for web server prediction when a model has been trained only on one group.

    \subsection{Future work}
    Besides analyzing and solving the aforementioned limitations of this work, we have identified potentially other interesting topics for future research.

    \subsubsection{Including the complete HTTP response headers}
    Perhaps the most obvious unit of work: re-run this experiment, but including the complete headers and with longer training times.
    In principle, no major architectural changes are required except for more disk space and computational power.

    \subsubsection{Universal test cases for fingerprinting independent of port and protocol}
        We limited ourselves to samples which responded on both protocols.
        Our system will need to be adapted to work with both protocols individually to make the system universally usable to detect any listening web server.
        This can be done by simply using the same test cases for each port and then creating two separate samples for each domain.
        However, this will result in less data per sample, and it is unknown what impact this will have.
        Especially considering that the HTTPS/443 responses were generally more powerful predictors than the HTTP/80 responses.

    \subsubsection{Careful test case design}
    Although we currently have a workable amount of diverse test cases, looking at the feature importances some should be tweaked or dropped altogether as they are redundant.
    It would be worthwhile to construct a well thought out list of tests.

    \subsubsection{Analyze Transformer attentions}
    \citet{gniewkowski2021http2vec} analyzed RoBERTa's attentions to determine which tokens were important for anomaly detection.
    This would be an interesting method to detect meaningful elements for fingerprinting as well.

    \subsubsection{Improve embedding quality}
    As was mentioned, the quality of the embeddings can be improved through a supervised or unsupervised semantic similarity task.
    This would constitute an improvement over our use of raw, pooled hidden states.

    \subsubsection{Approach the task as a sequence-to-sequence or multilabel problem}
    This would be an interesting alternative to classification, particularly as a solution to class sparsity.
    Transformers are typically implemented as encoder-decoder models, and would be suitable for this.

    \subsubsection{Use template generation to cluster the target classes}
    Implementing template generation as proposed by \citet{mizuno2018detecting} and \citet{niu2019using} would be an extremely promising improvement over our naive pre-processing of target classes.
    It could potentially solve the problem of class sparsity without having to drop any samples.

    \subsubsection{Perform exhaustive statistical data analysis and identify interesting patterns}
    As mentioned, we have not performed elaborate data analysis.
    It is without doubt that further statistical analysis would reveal possible flaws in our method and interesting patterns in the World Wide Web.

    \subsubsection{Investigate the feasibility of fine-tuning and transfer learning}
    It is possible to create verifiably correct samples of new software releases on various platforms.
    These could be used as strongly weighted samples for fine-tuning or transfer learning, to improve the reliability of correctness of the model.

    \subsubsection{Determine what constitutes a poor sample}
    For this experiment, we used arbitrary thresholds to determine whether a sample was valid or not.
    This could be refined into a more concrete evidence-based guideline.

    \subsubsection{Compare various tokenizers, Transformer models and hyperparameters}
    We have only tried the default RoBERTa model with its BPE tokenizer as an encoder.
    Future work could determine which model, tokenization method, and which hyperparameters are suitable for this fingerprinting task.

\subsubsection{Compare categorical encoding with Transformer encoding}
    In this work, we have not determined whether Transformer-encoding a single status line has any meaningful difference over categorically encoding it.
    Re-running this experiment, but with categorical encoding of the HTTP status lines could help determine this.
    Additionally, it would be interesting to analyze the effect of extending the list of status lines by artificially generating possible status lines in accordance with the grammar of the protocol.

    \subsubsection{Investigate the cause of high classification performance}
    We have not determined the root cause of our high classification performance.
    We wonder what exactly constitutes the fingerprints of e.g., some minor versions.
    Do some individual versions really have very specific fingerprints (and if so: which?), or is the data or method flawed somehow?

    \subsubsection{Analyze the performance of other classification algorithms}
    This experiment was limited to a feed forward network and Random Forest as classifiers.
    In the background literature, other algorithms such as XGBoost were commonly used.
    It would be worthwhile to analyze the performance differences between various classification algorithms.

    \subsubsection{Analyze the performance between dimensionality reduction methods}
    We have used principal component analysis to reduce the dimensionality of our test response data.
    However, there are other ways to achieve fewer dimensions, such as directly extending and training a BERT-like model with an additional final hidden layer of reduced dimensionality.
    With that method, lower-dimensional embeddings can be generated directly by the model.

    \subsubsection{Compare our performance with traditional rule-based systems}
    Although we would have liked to compare our results with existing rule-based systems as well, this is a non-trivial endeavour.
    These existing tools perform their classification on live systems and cannot work directly on downloaded data.
    We would have to run a tool on all domains, store the predicted class, and then compare the findings with ours.
    Critically, many of these rule-based systems actively make use of the \texttt{Server:} header to make their estimation.
    It is likely that source code would have to be modified in order to disable that.
    Comparing these systems would, however, be crucial in determining the direction of future fingerprinting research.
    Is our method a universal improvement? Or are the existing systems better in some regard?

	\nocite{10.5555/1538595, lee2001hmap}

	\bibliographystyle{acl_natbib}
	\bibliography{custom}

\begin{thebibliography}{37}
\expandafter\ifx\csname natexlab\endcsname\relax\def\natexlab#1{#1}\fi

\bibitem[{Agarap(2018)}]{https://doi.org/10.48550/arxiv.1803.08375}
Abien~Fred Agarap. 2018.
\newblock \href {https://doi.org/10.48550/ARXIV.1803.08375} {Deep learning using rectified linear units (relu)}.

\bibitem[{Al-Hakimi and Bax(2021)}]{alhunting}
Shadi Al-Hakimi and Freek Bax. 2021.
\newblock \href {https://rp.os3.nl/2020-2021/p54/report.pdf} {Hunting for malicious infrastructure using big data}.
\newblock \emph{SNE Master Research Projects 2020 - 2021, University of Amsterdam}.

\bibitem[{Barnes et~al.(2019)Barnes, Hoffman-Andrews, McCarney, and Kasten}]{rfc8555}
Richard Barnes, Jacob Hoffman-Andrews, Daniel McCarney, and James Kasten. 2019.
\newblock \href {https://doi.org/10.17487/RFC8555} {{Automatic Certificate Management Environment (ACME)}}.
\newblock RFC 8555.

\bibitem[{Belshe et~al.(2015)Belshe, Peon, and Thomson}]{rfc7540}
Mike Belshe, Roberto Peon, and Martin Thomson. 2015.
\newblock \href {https://doi.org/10.17487/RFC7540} {{Hypertext Transfer Protocol Version 2 (HTTP/2)}}.
\newblock RFC 7540.

\bibitem[{Beltagy et~al.(2020)Beltagy, Peters, and Cohan}]{beltagy2020longformer}
Iz~Beltagy, Matthew~E. Peters, and Arman Cohan. 2020.
\newblock \href {https://doi.org/10.48550/ARXIV.2004.05150} {Longformer: The long-document transformer}.

\bibitem[{Berners-Lee et~al.(1996)Berners-Lee, Fielding, and Frystyk}]{10.17487/RFC1945}
T.~Berners-Lee, R.~Fielding, and H.~Frystyk. 1996.
\newblock \href {https://www.rfc-editor.org/info/rfc1945} {{Hypertext Transfer Protocol -- HTTP/1.0}}.
\newblock RFC 1945.

\bibitem[{Book et~al.(2013)Book, Witick, and Wallach}]{book2013automated}
Theodore Book, Martha Witick, and Dan~S. Wallach. 2013.
\newblock \href {https://doi.org/10.48550/ARXIV.1305.0245} {Automated generation of web server fingerprints}.

\bibitem[{Devlin et~al.(2019)Devlin, Chang, Lee, and Toutanova}]{devlin-etal-2019-bert}
Jacob Devlin, Ming-Wei Chang, Kenton Lee, and Kristina Toutanova. 2019.
\newblock \href {https://doi.org/10.18653/v1/N19-1423} {{BERT}: Pre-training of deep bidirectional transformers for language understanding}.
\newblock In \emph{Proceedings of the 2019 Conference of the North {A}merican Chapter of the Association for Computational Linguistics: Human Language Technologies, Volume 1 (Long and Short Papers)}, pages 4171--4186, Minneapolis, Minnesota. Association for Computational Linguistics.

\bibitem[{Fielding et~al.(1997)Fielding, Nielsen, Mogul, Gettys, and Berners-Lee}]{rfc2068}
Roy~T. Fielding, Henrik Nielsen, Jeffrey Mogul, Jim Gettys, and Tim Berners-Lee. 1997.
\newblock \href {https://doi.org/10.17487/RFC2068} {{Hypertext Transfer Protocol -- HTTP/1.1}}.
\newblock RFC 2068.

\bibitem[{Gniewkowski et~al.(2021)Gniewkowski, Maciejewski, Surmacz, and Walentynowicz}]{gniewkowski2021http2vec}
Mateusz Gniewkowski, Henryk Maciejewski, Tomasz~R. Surmacz, and Wiktor Walentynowicz. 2021.
\newblock \href {https://doi.org/10.48550/ARXIV.2108.01763} {Http2vec: Embedding of http requests for detection of anomalous traffic}.

\bibitem[{Guliyev and Ismailov(2018)}]{GULIYEV2018262}
Namig~J. Guliyev and Vugar~E. Ismailov. 2018.
\newblock \href {https://doi.org/https://doi.org/10.1016/j.neucom.2018.07.075} {Approximation capability of two hidden layer feedforward neural networks with fixed weights}.
\newblock \emph{Neurocomputing}, 316:262--269.

\bibitem[{Heaton(2008)}]{10.5555/1502373}
Jeff Heaton. 2008.
\newblock \emph{Introduction to Neural Networks for Java, 2nd Edition}, 2nd edition.
\newblock Heaton Research, Inc.

\bibitem[{Huang et~al.(2015)Huang, Xia, Sun, and Xue}]{huang2015analyzing}
Zhen Huang, Chunhe Xia, Bo~Sun, and Hui Xue. 2015.
\newblock \href {https://doi.org/10.1109/ICSESS.2015.7339231} {Analyzing and summarizing the web server detection technology based on http}.
\newblock In \emph{2015 6th IEEE International Conference on Software Engineering and Service Science (ICSESS)}, pages 1042--1045.

\bibitem[{Kingma and Ba(2014)}]{https://doi.org/10.48550/arxiv.1412.6980}
Diederik~P. Kingma and Jimmy Ba. 2014.
\newblock \href {https://doi.org/10.48550/ARXIV.1412.6980} {Adam: A method for stochastic optimization}.

\bibitem[{Laughter et~al.(2021)Laughter, Omari, Szczurek, and Perry}]{laughter2020detection}
Ashley Laughter, Safwan Omari, Piotr Szczurek, and Jason Perry. 2021.
\newblock Detection of malicious http requests using header and url features.
\newblock In \emph{Proceedings of the Future Technologies Conference (FTC) 2020, Volume 2}, pages 449--468, Cham. Springer International Publishing.

\bibitem[{Lee et~al.(2002)Lee, Rowe, Ko, and Levitt}]{lee2002detecting}
D.~Lee, J.~Rowe, C.~Ko, and K.~Levitt. 2002.
\newblock \href {https://doi.org/10.1109/CSAC.2002.1176304} {Detecting and defending against web-server fingerprinting}.
\newblock In \emph{18th Annual Computer Security Applications Conference, 2002. Proceedings.}, pages 321--330.

\bibitem[{Lee(2001)}]{lee2001hmap}
Dustin~William Lee. 2001.
\newblock \href {https://seclab.cs.ucdavis.edu/papers/hmap-thesis.pdf} {Hmap: A technique and tool for remote identification of http servers}.

\bibitem[{Li et~al.(2020)Li, Zhang, and Wei}]{li2020weighted}
Jieling Li, Hao Zhang, and Zhiqiang Wei. 2020.
\newblock \href {https://doi.org/10.1109/ACCESS.2020.3013849} {The weighted word2vec paragraph vectors for anomaly detection over http traffic}.
\newblock \emph{IEEE Access}, 8:141787--141798.

\bibitem[{Liu et~al.(2019)Liu, Ott, Goyal, Du, Joshi, Chen, Levy, Lewis, Zettlemoyer, and Stoyanov}]{https://doi.org/10.48550/arxiv.1907.11692}
Yinhan Liu, Myle Ott, Naman Goyal, Jingfei Du, Mandar Joshi, Danqi Chen, Omer Levy, Mike Lewis, Luke Zettlemoyer, and Veselin Stoyanov. 2019.
\newblock \href {https://doi.org/10.48550/ARXIV.1907.11692} {Roberta: A robustly optimized bert pretraining approach}.

\bibitem[{Lyon(2009)}]{10.5555/1538595}
Gordon~Fyodor Lyon. 2009.
\newblock \emph{Nmap Network Scanning: The Official Nmap Project Guide to Network Discovery and Security Scanning}.
\newblock Insecure, Sunnyvale, CA, USA.

\bibitem[{Mizuno et~al.(2018)Mizuno, Hatada, Mori, and Goto}]{mizuno2018detecting}
Sho Mizuno, Mitsuhiro Hatada, Tatsuya Mori, and Shigeki Goto. 2018.
\newblock \href {https://doi.org/10.1587/transinf.2017EDP7294} {Detecting malware-infected devices using the http header patterns}.
\newblock \emph{IEICE Transactions on Information and Systems}, E101.D(5):1370--1379.

\bibitem[{Niakanlahiji et~al.(2018)Niakanlahiji, Chu, and Al-Shaer}]{niakanlahiji2018phishmon}
Amirreza Niakanlahiji, Bei-Tseng Chu, and Ehab Al-Shaer. 2018.
\newblock \href {https://doi.org/10.1109/ISI.2018.8587410} {Phishmon: A machine learning framework for detecting phishing webpages}.
\newblock In \emph{2018 IEEE International Conference on Intelligence and Security Informatics (ISI)}, pages 220--225.

\bibitem[{Niu et~al.(2019)Niu, Li, Zhang, Hu, Jiang, and Wu}]{niu2019using}
Weina Niu, Ting Li, Xiaosong Zhang, Teng Hu, Tianyu Jiang, and Heng Wu. 2019.
\newblock \href {https://doi.org/10.1155/2019/2182615} {Using xgboost to discover infected hosts based on http traffic}.
\newblock \emph{Security and Communication Networks}, 2019:2182615.

\bibitem[{Pochat et~al.(2019)Pochat, Goethem, Tajalizadehkhoob, Korczynski, and Joosen}]{pochat2018tranco}
Victor~Le Pochat, Tom~Van Goethem, Samaneh Tajalizadehkhoob, Maciej Korczynski, and Wouter Joosen. 2019.
\newblock \href {https://doi.org/10.14722/ndss.2019.23386} {Tranco: A research-oriented top sites ranking hardened against manipulation}.
\newblock In \emph{Proceedings 2019 Network and Distributed System Security Symposium}. Internet Society.

\bibitem[{Reimers and Gurevych(2019)}]{reimers-2019-sentence-bert}
Nils Reimers and Iryna Gurevych. 2019.
\newblock \href {http://arxiv.org/abs/1908.10084} {Sentence-bert: Sentence embeddings using siamese bert-networks}.
\newblock In \emph{Proceedings of the 2019 Conference on Empirical Methods in Natural Language Processing}. Association for Computational Linguistics.

\bibitem[{Ruef(2007)}]{ruef_2007}
Marc Ruef. 2007.
\newblock \href {https://www.computec.ch/projekte/httprecon/?s=documentation} {httprecon project - advanced web server fingerprinting}.

\bibitem[{Shah(2004)}]{shah_2004}
Saumil Shah. 2004.
\newblock \href {https://net-square.com/httprint_paper.html} {An introduction to http fingerprinting}.

\bibitem[{Srivastava et~al.(2014)Srivastava, Hinton, Krizhevsky, Sutskever, and Salakhutdinov}]{JMLR:v15:srivastava14a}
Nitish Srivastava, Geoffrey Hinton, Alex Krizhevsky, Ilya Sutskever, and Ruslan Salakhutdinov. 2014.
\newblock \href {http://jmlr.org/papers/v15/srivastava14a.html} {Dropout: A simple way to prevent neural networks from overfitting}.
\newblock \emph{Journal of Machine Learning Research}, 15(56):1929--1958.

\bibitem[{{van der Maaten} and Hinton(2008)}]{7b54165e73a3424b8820136bcf61ca89}
L.J.P. {van der Maaten} and G.E. Hinton. 2008.
\newblock Visualizing high-dimensional data using t-sne.
\newblock \emph{Journal of Machine Learning Research}, 9(nov):2579--2605.
\newblock Pagination: 27.

\bibitem[{Vaswani et~al.(2017)Vaswani, Shazeer, Parmar, Uszkoreit, Jones, Gomez, Kaiser, and Polosukhin}]{vaswani}
Ashish Vaswani, Noam Shazeer, Niki Parmar, Jakob Uszkoreit, Llion Jones, Aidan~N. Gomez, \L{}ukasz Kaiser, and Illia Polosukhin. 2017.
\newblock Attention is all you need.
\newblock In \emph{Proceedings of the 31st International Conference on Neural Information Processing Systems}, NIPS'17, page 6000–6010, Red Hook, NY, USA. Curran Associates Inc.

\bibitem[{Wang et~al.(2021)Wang, Reimers, and Gurevych}]{wang-2021-TSDAE}
Kexin Wang, Nils Reimers, and Iryna Gurevych. 2021.
\newblock \href {https://arxiv.org/abs/2104.06979} {Tsdae: Using transformer-based sequential denoising auto-encoder for unsupervised sentence embedding learning}.
\newblock \emph{arXiv preprint arXiv:2104.06979}.

\bibitem[{Wolf et~al.(2019)Wolf, Debut, Sanh, Chaumond, Delangue, Moi, Cistac, Rault, Louf, Funtowicz, and Brew}]{DBLP:journals/corr/abs-1910-03771}
Thomas Wolf, Lysandre Debut, Victor Sanh, Julien Chaumond, Clement Delangue, Anthony Moi, Pierric Cistac, Tim Rault, R{\'{e}}mi Louf, Morgan Funtowicz, and Jamie Brew. 2019.
\newblock \href {http://arxiv.org/abs/1910.03771} {Huggingface's transformers: State-of-the-art natural language processing}.
\newblock \emph{CoRR}, abs/1910.03771.

\bibitem[{Yang et~al.(2010)Yang, Hu, Zhang, Huo, and Zhao}]{yang2010improving}
Ke-xin Yang, Liang Hu, Ning Zhang, Yan-mei Huo, and Kuo Zhao. 2010.
\newblock \href {https://doi.org/10.1109/FCST.2010.91} {Improving the defence against web server fingerprinting by eliminating compliance variation}.
\newblock In \emph{2010 Fifth International Conference on Frontier of Computer Science and Technology}, pages 227--232.

\bibitem[{Yu et~al.(2018)Yu, Liu, Yan, Li, and Guan}]{yu2018attention}
Yuqi Yu, Guannan Liu, Hanbing Yan, Hong Li, and Hongchao Guan. 2018.
\newblock \href {https://doi.org/10.1109/ICSSSM.2018.8465034} {Attention-based bi-lstm model for anomalous http traffic detection}.
\newblock In \emph{2018 15th International Conference on Service Systems and Service Management (ICSSSM)}, pages 1--6.

\bibitem[{Yuan et~al.(2018{\natexlab{a}})Yuan, Chen, Li, Yang, and Liu}]{yuan2018detecting}
Huaping Yuan, Xu~Chen, Yukun Li, Zbenguo Yang, and Wenyin Liu. 2018{\natexlab{a}}.
\newblock \href {https://doi.org/10.1109/ICPR.2018.8546262} {Detecting phishing websites and targets based on urls and webpage links}.
\newblock In \emph{2018 24th International Conference on Pattern Recognition (ICPR)}, pages 3669--3674.

\bibitem[{Yuan et~al.(2018{\natexlab{b}})Yuan, Yang, Chen, Li, and Liu}]{yuan2018url2vec}
Huaping Yuan, Zhenguo Yang, Xu~Chen, Yukun Li, and Wenyin Liu. 2018{\natexlab{b}}.
\newblock \href {https://doi.org/10.1109/BDCloud.2018.00050} {Url2vec: Url modeling with character embeddings for fast and accurate phishing website detection}.
\newblock In \emph{2018 IEEE Intl Conf on Parallel \& Distributed Processing with Applications, Ubiquitous Computing \& Communications, Big Data \& Cloud Computing, Social Computing \& Networking, Sustainable Computing \& Communications (ISPA/IUCC/BDCloud/SocialCom/SustainCom)}, pages 265--272.

\bibitem[{Zaheer et~al.(2020)Zaheer, Guruganesh, Dubey, Ainslie, Alberti, Ontanon, Pham, Ravula, Wang, Yang, and Ahmed}]{bigbird2020}
Manzil Zaheer, Guru~Prashanth Guruganesh, Avinava Dubey, Joshua Ainslie, Chris Alberti, Santiago Ontanon, Philip~Minh Pham, Anirudh Ravula, Qifan Wang, Li~Yang, and Amr Mahmoud El~Houssieny Ahmed. 2020.
\newblock \href {https://proceedings.neurips.cc/paper/2020/file/c8512d142a2d849725f31a9a7a361ab9-Paper.pdf} {Big bird: Transformers for longer sequences}.

\end{thebibliography}

	\appendix

	\section{Appendices}

	\subsection{Source code and dataset}

	The code of this project can be found on GitHub\footnote{\url{https://github.com/Darwinkel/bachelor-thesis-information-science}}, and is licensed under the GNU GPLv3. The data that was
    collected and processed is licensed under the Attribution-ShareAlike 4.0 International (CC BY-SA 4.0) license and
    can be found on GitHub as well.
    Any interested party is free to contact me about the project itself and about potential future work.

    \subsection{Acknowledgements}
	This work would not have been possible without the support and input from my friends and colleagues. In particular, I would like to thank Malvina Nissim and Frank Tsiwah from the University of Groningen for the academic supervision of this work.

    \subsection{Ethics statement}
    In principle, the techniques outlined in this work may be used by malicious actors to find security holes in
    existing server systems. It could be said that this work is, in fact, detrimental to the security of the
    World Wide Web and should not be released to the public. It is our hope that the outcome of this research will
    stimulate:

    \begin{itemize}
        \item the Internet Engineering Task Force and the World Wide Web Consortium to eliminate unnecessary
        ambiguity from the HTTP specification;
        \item web server vendors to synchronize their implementations of the HTTP specification;
        \item system administrators to keep their systems updated, and configure them not to reveal version information.
    \end{itemize}

    Furthermore, all HTTP requests made on behalf of this experiment were designed to be non-destructive. Unless a
    web server was seriously misconfigured, no websites were harmed in the process.

    \subsection{Examples of HTTP response headers}\label{sec:responseexamples}

    \begin{figure}
        \centering
        \begin{verbatim}
HTTP/1.1 200 OK
Date: Thu, 05 Sep 2019 17:42:39 GMT
Server: Apache/2.4.41 (Unix)
Last-Modified: Thu, 05 Sep 2019 17:40:42 GMT
ETag: "75-591d1d21b6167"
Accept-Ranges: bytes
Content-Length: 117
Connection: close
Content-Type: text/html
        \end{verbatim}

        \caption{An example of a HTTP response header by an Apache web server. Courtesy of the Open Web
        Application Security Project.\label{figure:apache_example}}
    \end{figure}

    \begin{figure}
        \centering
        \begin{verbatim}
HTTP/1.1 200 OK
Server: nginx/1.17.3
Date: Thu, 05 Sep 2019 17:50:24 GMT
Content-Type: text/html
Content-Length: 117
Last-Modified: Thu, 05 Sep 2019 17:40:42 GMT
Connection: close
ETag: "5d71489a-75"
Accept-Ranges: bytes
        \end{verbatim}

        \caption{An example of a HTTP response header by an nginx web server. Courtesy of the Open Web
        Application Security Project.\label{figure:nginx_example}}
    \end{figure}

    \begin{figure}
        \centering
        \begin{verbatim}
HTTP/1.0 200 OK
Content-Type: text/html
Accept-Ranges: bytes
ETag: "4192788355"
Last-Modified: Thu, 05 Sep 2019 17:40:42 GMT
Content-Length: 117
Connection: close
Date: Thu, 05 Sep 2019 17:57:57 GMT
Server: lighttpd/1.4.54
        \end{verbatim}

        \caption{An example of a HTTP response header by a lighthttpd web server. Courtesy of the Open Web
        Application Security Project.\label{figure:lighthttpd_example}}
    \end{figure}

	    \begin{figure*}
			\centering
	\begin{verbatim}
HTTP/1.1 404 no soup for you!
HTTP/1.1 403 YOU SHALL NOT PASS
HTTP/1.1 418 I'm a teapot (RFC 2324)
HTTP/1.1 999 No Hacking CambridgeSoft
HTTP/1.1 301 Go west, life is peaceful there
HTTP/1.1 505 You gone bongers mate?
HTTP/1.1 200 I'm sorry, Dave. I'm afraid I can't work without a host header.
HTTP/1.1 403 Naughty, not nice!
HTTP/1.1 302 - Fake found, kill Bill Gates and IE 4+ with it's 404 page
\end{verbatim}
	\caption{Examples of non-standard HTTP status lines.\label{sec:interestingstatuslines}}
	\end{figure*}

    \newpage

    \subsection{Test case descriptions}\label{sec:testcases}

    Note that \texttt{\$hostname} will be replaced by the domain name, e.g., \texttt{www.google.com}.
    Also note that the provided \texttt{User-Agent:} is forged.
    We use the user agent string of a common web browser to reduce the
    chances that our requests are blocked by spam filters.
	In the below descriptions, \texttt{(...)} should be expanded to \texttt{(X11; Linux x86\_64; rv:105.0) Gecko/20100101 Firefox/105.0}

    \paragraph{Test case 0}
    The result after following a maximum of 3 redirects to the same port and protocol.
    Since we collect this information regardless of circumstances, it makes sense to use as a baseline result for a
    valid request. Suggested by \citet{lee2002detecting}, \citet{shah_2004}, \citet{ruef_2007}, and
    \citet{book2013automated}.

    \begin{verbatim}
HEAD / HTTP/1.1\r\n
Host: $hostname\r\n
User-Agent: Mozilla/5.0 (...)\r\n
Accept: */*\r\n
\r\n
    \end{verbatim}

    \paragraph{Test case 1}
    An HTTP/2 request where we try to negotiate support. The specification recommends, but not requires, `101
    Switching Protocols`. Original.\\

    HTTP/80 (h2c)
    \begin{verbatim}
HEAD / HTTP/1.1\r\n
Host: $hostname\r\n
User-Agent: Mozilla/5.0 (...)\r\n
Accept: */*\r\n
Connection: Upgrade, HTTP2-Settings\r\n
Upgrade: h2c\r\n
HTTP2-Settings: AAMAAABkAAQCAAAAAAIAAAAA\r\n
\r\n
    \end{verbatim}

    HTTPS/443 (h2)
    \begin{verbatim}
HEAD / HTTP/2\r\n
Host: $hostname\r\n
user-agent: Mozilla/5.0 (...)\r\n
accept: */*\r\n
\r\n
    \end{verbatim}

    \paragraph{Test case 2}
    An HTTP/2 request where we assume support. Original.\\

    HTTP/80 (h2c)
    \begin{verbatim}
HEAD / HTTP/2\r\n
Host: $hostname\r\n
user-agent: Mozilla/5.0 (...)\r\n
accept: */*\r\n
\r\n
    \end{verbatim}

    HTTPS/443 (h2)
    \begin{verbatim}
HEAD / HTTP/2\r\n
Host: $hostname\r\n
user-agent: Mozilla/5.0 (...)\r\n
accept: */*\r\n
\r\n
    \end{verbatim}

    \paragraph{Test case 3}
    A regular GET request to the webserver root, according to the 1990 HTTP/0.9 specification. HTTP/0.9 has been
    completely deprecated, and many web servers reply with an explicit HTTP/1.1 response. If software supports this, then the header reply should
    be completely empty. The specification recommends `426 Upgrade Required` if a compliant response will not be
    given. Original.

    \begin{verbatim}
GET /\r\n
\r\n
    \end{verbatim}

    \paragraph{Test case 4}
    A regular GET request to the webserver root, according to the 1996 HTTP/1.0 specification. HTTP/1.0 is almost
    never used nowadays, and many web servers reply with an explicit HTTP/1.1 response. Some, however, do give an
    HTTP/1.0 compliant response. The specification recommends `426 Upgrade Required` if a compliant response will not
    be given. Original.

    \begin{verbatim}
GET / HTTP/1.0\r\n
User-Agent: Mozilla/5.0 (...)\r\n
\r\n
    \end{verbatim}

    \paragraph{Test case 5}
    A regular GET request to the webserver root, according to the 1996 HTTP/1.0 specification, but including the
    \texttt{Host:} header for website disambiguation. Original.

    \begin{verbatim}
GET / HTTP/1.0\r\n
User-Agent: Mozilla/5.0 (...)\r\n
Host: $hostname\r\n
\r\n
    \end{verbatim}

    \paragraph{Test case 6}
    A regular HEAD request to the webserver root, but with \texttt{\textbackslash n} line endings. This is not
    compliant with the HTTP specification, but it is the standard terminator token for Unix and Unix-like operating
    systems. Original.

    \begin{verbatim}
HEAD / HTTP/1.1\n
User-Agent: Mozilla/5.0 (...)\n
Host: $hostname\n
\n
    \end{verbatim}

    \paragraph{Test case 7}
    A regular HEAD request to the webserver root, but with \texttt{\textbackslash r} line endings. This is not
    compliant with the HTTP specification, but is used as the standard terminator token by some legacy operating
    systems. Original.

    \begin{verbatim}
HEAD / HTTP/1.1\r
User-Agent: Mozilla/5.0 (...)\r
Host: $hostname\r
\r
    \end{verbatim}

    \paragraph{Test case 8}
    A regular HEAD request to the webserver root, but specifying the non-existent HTTP/4.0 protocol. The
    specification recommends, but not requires, the `505 HTTP Version Not Supported` status. Suggested by
    \citet{lee2002detecting}, \citet{huang2015analyzing}, \citet{shah_2004}, and \citet{ruef_2007}.

    \begin{verbatim}
HEAD / HTTP/4.0\r\n
Host: $hostname\r\n
User-Agent: Mozilla/5.0 (...)\r\n
Accept: */*\r\n
\r\n
    \end{verbatim}

    \paragraph{Test case 9}
    A regular HEAD request to the webserver root, but specifying the non-existent SPOCK method. The specification
    recommends, but not requires, the `501 Not Implemented` status.  Suggested by
    \citet{lee2002detecting}, \citet{huang2015analyzing}, \citet{shah_2004}, and \citet{ruef_2007}.

    \begin{verbatim}
SPOCK / HTTP/1.1\r\n
Host: $hostname\r\n
User-Agent: Mozilla/5.0 (...)\r\n
Accept: */*\r\n
\r\n
    \end{verbatim}

    \paragraph{Test case 10}
    A HEAD request, that is case-insensitive. This is not compliant with the specification. Many variations on this
    exist (missing spaces, either method or protocol in lower case, misspellings etc.), but we have chosen to limit
    ourselves to this instance. Suggested by \citet{lee2002detecting} and \citet{huang2015analyzing}.

    \begin{verbatim}
HEad / httP/1.1\r\n
Host: $hostname\r\n
User-Agent: Mozilla/5.0 (...)\r\n
Accept: */*\r\n
\r\n
    \end{verbatim}

    \paragraph{Test case 11}
    An OPTIONS request to the webserver root. This is a request for metadata about what settings are supported by the
    server and the location. Suggested by \citet{book2013automated} and \citet{ruef_2007}.

    \begin{verbatim}
OPTIONS / HTTP/1.1\r\n
Host: $hostname\r\n
User-Agent: Mozilla/5.0 (...)\r\n
Accept: */*\r\n
\r\n
    \end{verbatim}

    \paragraph{Test case 12}
    A multipart GET request to the webserver root, requesting bytes 0-4 and 6-10 of the content. The specification
    recommends `206 Partial Content` for a success, and `416 Range Not Satisfiable` for a fail. Suggested
    by \citet{book2013automated}.

    \begin{verbatim}
HEAD / HTTP/1.1\r\n
Host: $hostname\r\n
User-Agent: Mozilla/5.0 (...)\r\n
Accept: */*\r\n
Range: bytes=0-4, 6-10\r\n
\r\n
    \end{verbatim}

    \paragraph{Test case 13}
    A conditional HEAD request, if content has been modified since a moment in the future. The specification
    recommends `304 Not Modified`. Suggested by \citet{book2013automated}.

    \begin{verbatim}
HEAD / HTTP/1.1\r\n
Host: $hostname\r\n
User-Agent: Mozilla/5.0 (...)\r\n
Accept: */*\r\n
If-Modified-Since: Wed, 21 Oct 2029 07:28:01 GMT\r\n
\r\n
    \end{verbatim}

    \paragraph{Test case 14}
    A conditional HEAD request, if content has not been modified since a moment in the distant past. The
    specification recommends `412 Precondition Failed`. Suggested by \citet{book2013automated}.

    \begin{verbatim}
HEAD / HTTP/1.1\r\n
Host: $hostname\r\n
User-Agent: Mozilla/5.0 (...)\r\n
Accept: */*\r\n
If-Unmodified-Since: Wed, 21 Oct 1985 07:28:01 GMT\r\n
\r\n
    \end{verbatim}

    \paragraph{Test case 15}
    A regular HEAD request to the webserver root, but we request that the content is returned as JavaScript or as a
    CSS stylesheet. It is extremely uncommon for the root to actually be a JavaScript or CSS file, and a secure web
    server should not naively comply, as this is may indicate a cross site scripting attack being in progress. The
    specification suggests `406 Not Acceptable` when refusing to deliver one of the acceptable formats.
    Suggested by \citet{book2013automated}.

    \begin{verbatim}
HEAD / HTTP/1.1\r\n
Host: $hostname\r\n
User-Agent: Mozilla/5.0 (...)\r\n
Accept: application/javascript;text/css;q=0.2\r\n
\r\n
    \end{verbatim}

    \subsection{Confusion matrices}\label{sec:cms}

                \begin{figure*}
        \centering
        \includegraphics[width=6in]{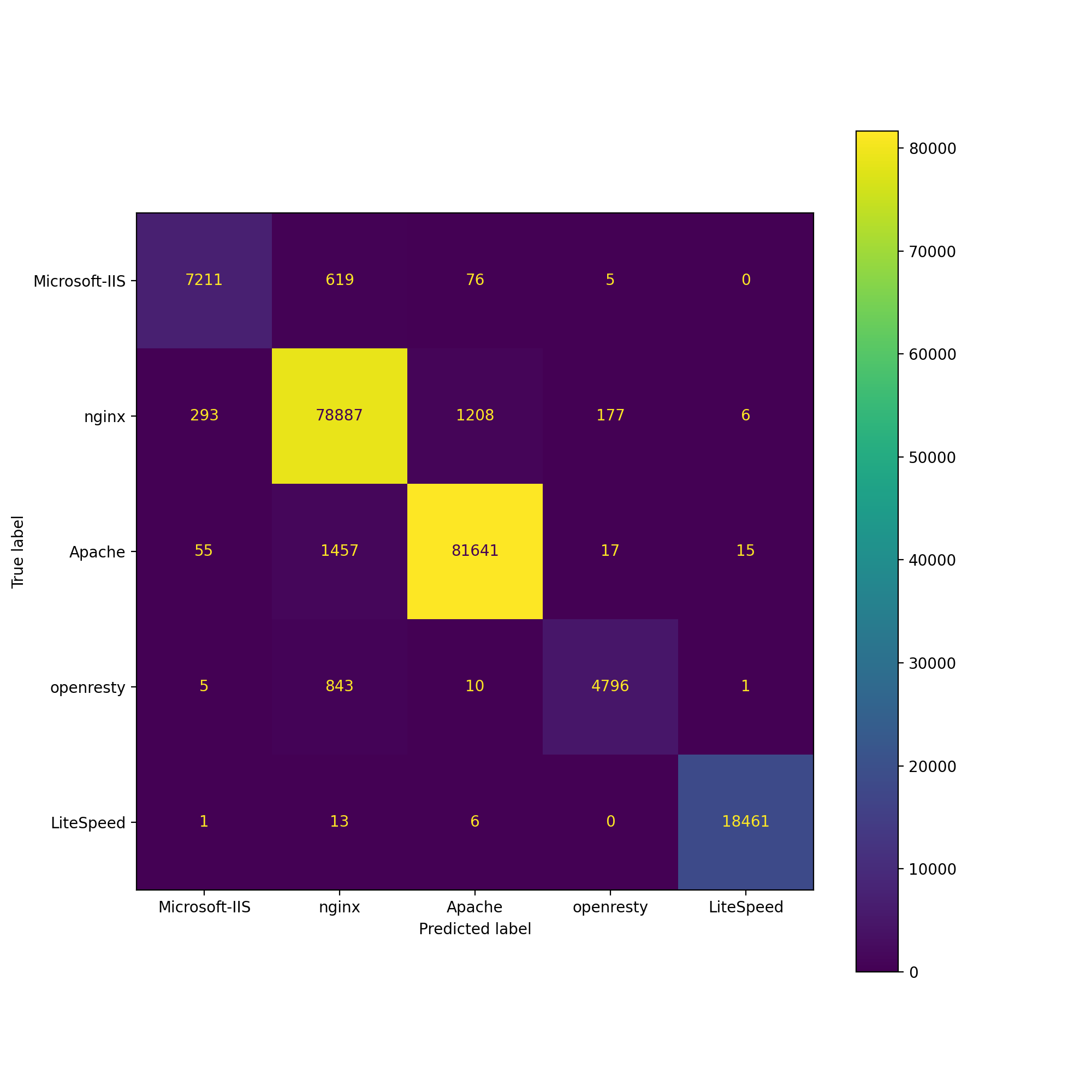}
        \caption{Confusion matrix of major type classification by a FNN.\label{figure:ffnn_majortypes_testset_abs}}
    \end{figure*}

                \begin{figure*}
        \centering
        \includegraphics[width=6in]{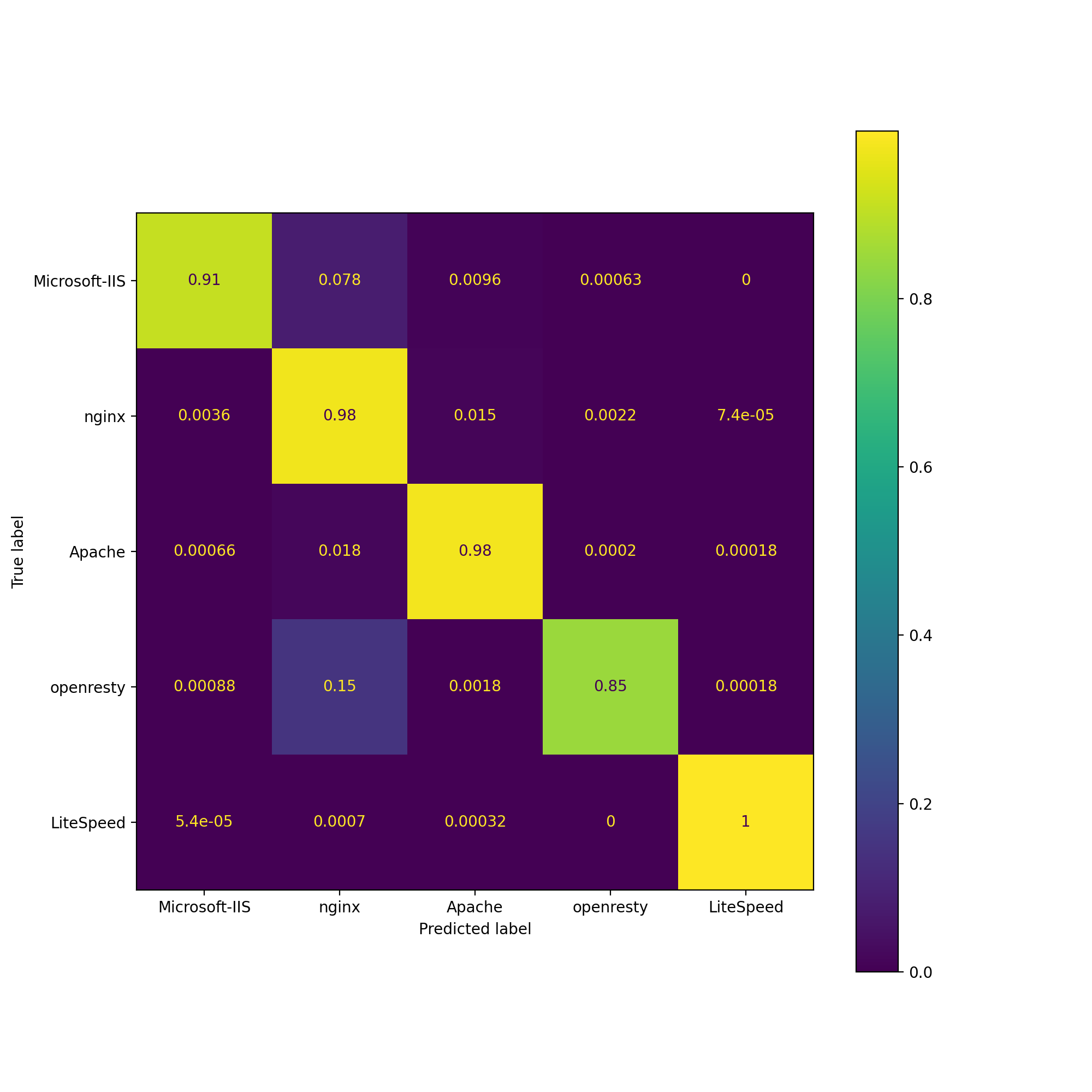}
        \caption{Confusion matrix of major type classification by a FNN, normalized row-wise by true labels.\label{figure:ffnn_majortypes_testset_normalized}}
    \end{figure*}

                \begin{figure*}
        \centering
        \includegraphics[width=6in]{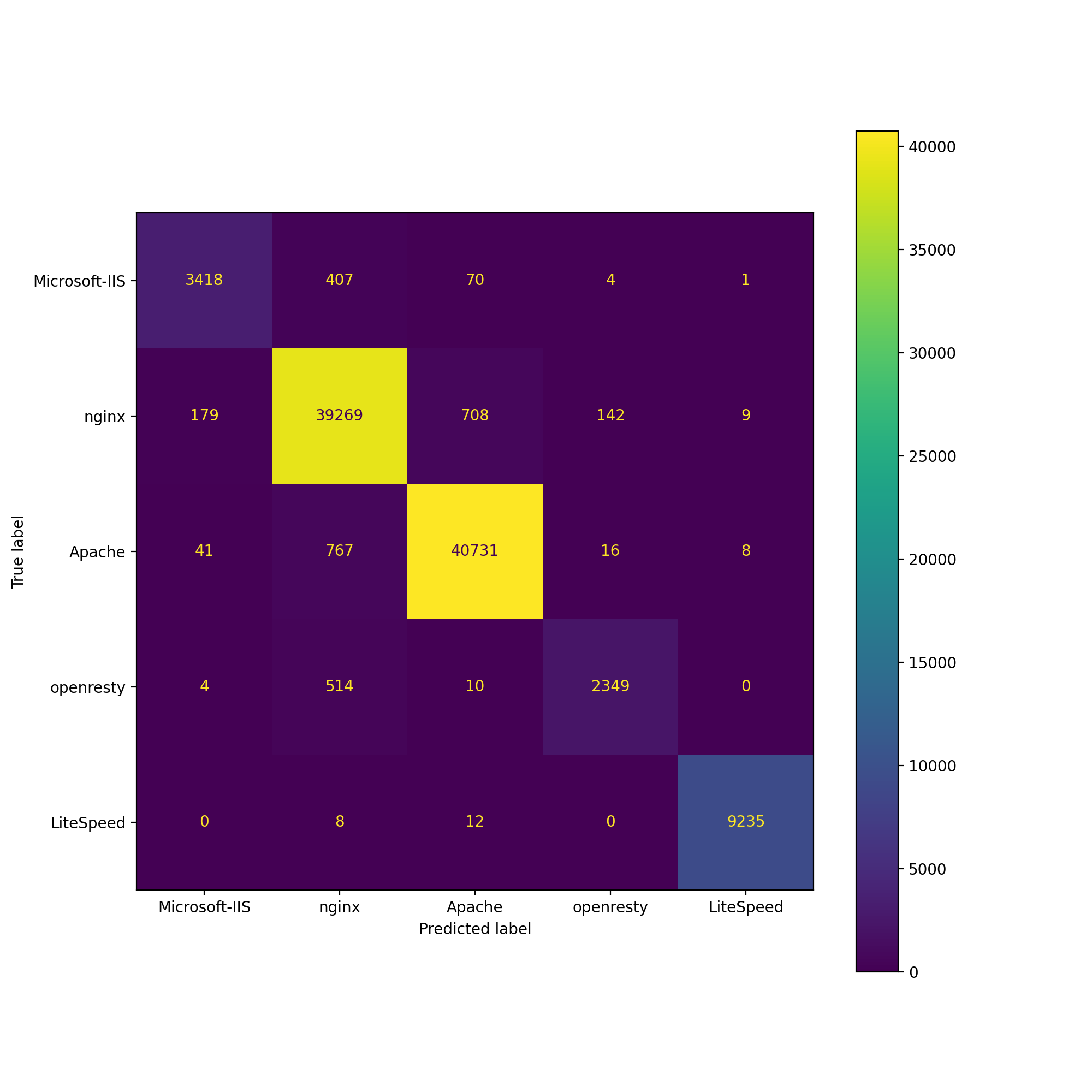}
        \caption{Confusion matrix of major type classification by a Random Forest.\label{figure:rf_majortypes_testset_abs}}
    \end{figure*}

                \begin{figure*}
        \centering
        \includegraphics[width=6in]{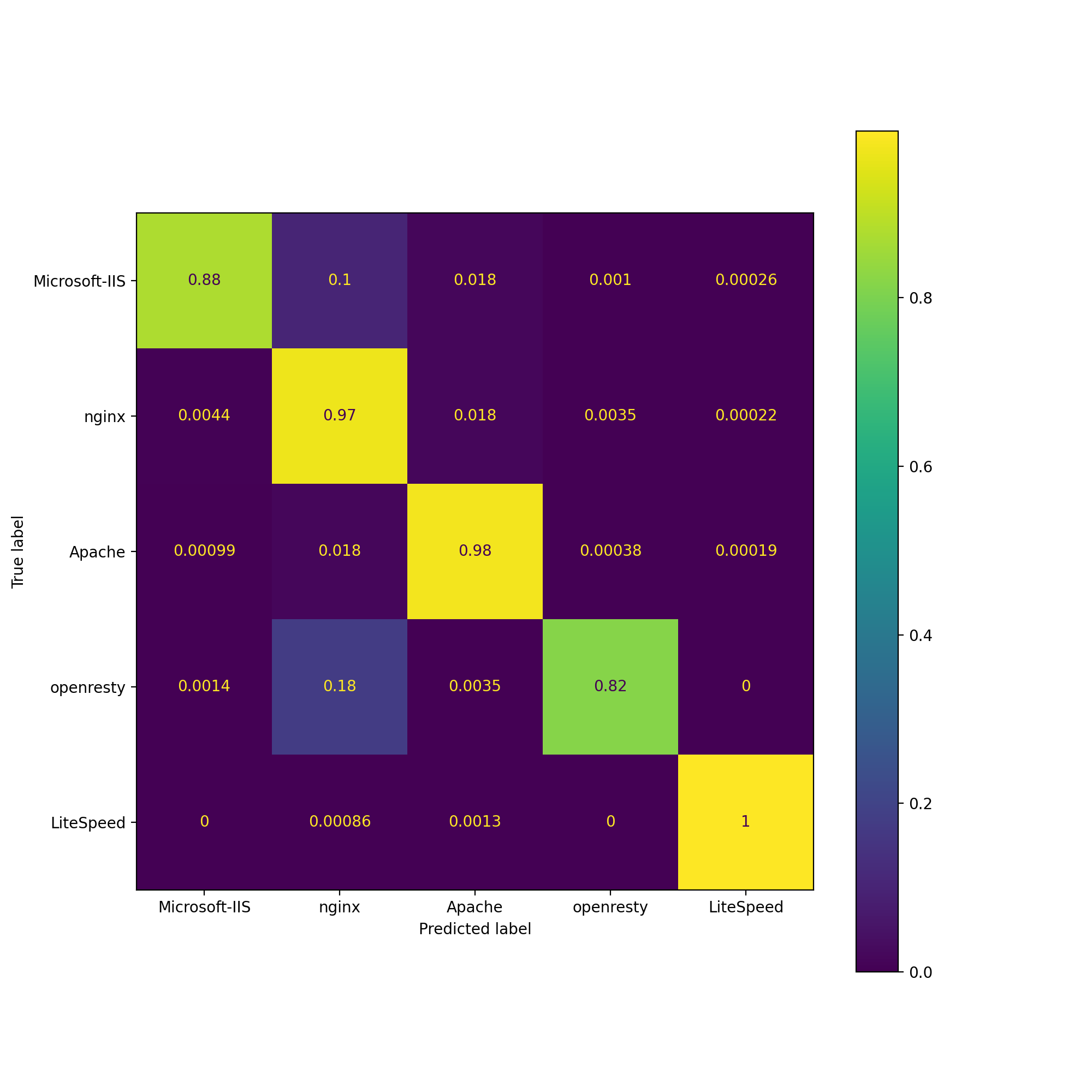}
        \caption{Confusion matrix of major type classification by a Random Forest, normalized row-wise by true labels.\label{figure:rf_majortypes_testset_normalized}}
    \end{figure*}

    \newpage

    \subsection{Classification reports}\label{sec:classificationreports}

\clearpage
\onecolumn
\begin{longtable}[c]{@{}llllr@{}}
\caption{Classification report for the FNN classifier with web server versions. All 347 classes are included in the calculations, but those with fewer than 3 occurrences are omitted from the table.}\label{tab:classification-report-ffnn-minortypes}\\
\toprule
                       & precision & recall & \textbf{f1-score} & support \\ \midrule
\endfirsthead

\multicolumn{5}{c}
{{\bfseries Table \thetable\ continued from previous page}} \\
\toprule
                       & precision & recall & \textbf{f1-score} & support \\ \midrule
\endhead
nginx/1.20.1           & 0.52      & 0.22   & 0.31              & 1216    \\
microsoft-iis/6.0      & 0.00      & 0.00   & 0.00              & 4       \\
nginx/1.15.0           & 0.00      & 0.00   & 0.00              & 19      \\
apache/2.4.35          & 0.67      & 0.11   & 0.18              & 19      \\
nginx/1.17.7           & 0.50      & 0.17   & 0.25              & 6       \\
nginx/1.12.0           & 0.56      & 0.23   & 0.33              & 43      \\
microsoft-iis/7.0      & 0.36      & 0.33   & 0.35              & 12      \\
apache/2.2             & 0.96      & 0.79   & 0.86              & 28      \\
nginx/1.12.1           & 0.83      & 0.27   & 0.41              & 124     \\
nginx/1.10.3           & 0.60      & 0.87   & 0.71              & 917     \\
nginx/1.14.0           & 0.33      & 0.10   & 0.16              & 1328    \\
nginx/1.10.2           & 0.17      & 0.09   & 0.11              & 92      \\
apache/2.4.26          & 0.00      & 0.00   & 0.00              & 8       \\
apache/2.4             & 0.95      & 0.66   & 0.78              & 161     \\
apache/2.4.17          & 0.40      & 0.20   & 0.27              & 10      \\
openresty/1.13.6.2     & 0.50      & 0.41   & 0.45              & 17      \\
nginx/1.19.3           & 0.67      & 0.14   & 0.24              & 28      \\
nginx/1.6.3            & 0.00      & 0.00   & 0.00              & 13      \\
nginx/1.13.0           & 0.00      & 0.00   & 0.00              & 7       \\
nginx/1.21.3           & 0.94      & 0.59   & 0.72              & 141     \\
apache/2.2.34          & 0.47      & 0.17   & 0.25              & 122     \\
nginx/1.19.0           & 0.57      & 0.48   & 0.52              & 44      \\
nginx/1.19.1           & 0.96      & 0.94   & 0.95              & 452     \\
nginx/1.22.1           & 0.61      & 0.20   & 0.30              & 56      \\
nginx/1.15.3           & 0.77      & 0.45   & 0.57              & 22      \\
apache/2.4.7           & 0.28      & 0.19   & 0.23              & 330     \\
nginx/1.6.1            & 0.00      & 0.00   & 0.00              & 3       \\
nginx/1.15.7           & 0.33      & 0.29   & 0.31              & 7       \\
apache/2.2.14          & 1.00      & 0.17   & 0.29              & 6       \\
nginx/1.4.7            & 0.20      & 0.25   & 0.22              & 8       \\
nginx/1.9.6            & 0.50      & 0.67   & 0.57              & 3       \\
nginx/1.6.0            & 0.17      & 0.08   & 0.11              & 12      \\
nginx/1.1.19           & 0.77      & 0.43   & 0.56              & 23      \\
nginx/1.13.3           & 0.00      & 0.00   & 0.00              & 26      \\
nginx/1.8.1            & 0.68      & 0.30   & 0.42              & 56      \\
nginx/1.11.10          & 0.00      & 0.00   & 0.00              & 4       \\
nginx/1.13.7           & 0.60      & 0.33   & 0.43              & 9       \\
apache/2.4.23          & 0.28      & 0.43   & 0.34              & 30      \\
openresty/1.15.8.2     & 0.40      & 0.40   & 0.40              & 42      \\
nginx/1.23.0           & 0.00      & 0.00   & 0.00              & 26      \\
apache/2.2.31          & 0.23      & 0.21   & 0.22              & 47      \\
nginx/1.21.5           & 0.00      & 0.00   & 0.00              & 21      \\
apache/2.4.27          & 0.00      & 0.00   & 0.00              & 43      \\
nginx/1.12.2           & 0.57      & 0.53   & 0.55              & 401     \\
nginx/1.13.9           & 1.00      & 0.17   & 0.29              & 6       \\
nginx/1.14.1           & 0.93      & 0.37   & 0.53              & 1444    \\
openresty/1.17.8.2     & 1.00      & 0.68   & 0.81              & 19      \\
nginx/1.11.5           & 0.67      & 0.40   & 0.50              & 5       \\
microsoft-iis/7.5      & 0.68      & 0.41   & 0.51              & 425     \\
nginx/1.21.1           & 0.55      & 0.32   & 0.40              & 69      \\
nginx/1.20.2           & 0.50      & 0.67   & 0.57              & 2090    \\
apache/2.4.34          & 1.00      & 0.02   & 0.05              & 43      \\
nginx/1.13.5           & 1.00      & 0.33   & 0.50              & 3       \\
apache/2.2.17          & 0.77      & 0.71   & 0.74              & 14      \\
microsoft-iis/8.0      & 0.71      & 0.06   & 0.10              & 212     \\
openresty/1.15.8.3     & 0.87      & 0.89   & 0.88              & 45      \\
nginx/1.23.1           & 0.40      & 0.16   & 0.23              & 151     \\
openresty/1.19.3.2     & 0.88      & 0.60   & 0.71              & 25      \\
nginx/1.15.11          & 0.67      & 0.22   & 0.33              & 9       \\
apache/2.4.6           & 0.65      & 0.95   & 0.77              & 2457    \\
apache/2.4.46          & 0.57      & 0.08   & 0.13              & 407     \\
nginx/1.15.9           & 0.33      & 0.05   & 0.08              & 21      \\
nginx/1.2.1            & 0.48      & 0.62   & 0.54              & 34      \\
nginx/1.11.3           & 0.40      & 0.33   & 0.36              & 18      \\
nginx/1.17.4           & 0.00      & 0.00   & 0.00              & 10      \\
nginx/1.20.0           & 0.48      & 0.28   & 0.36              & 176     \\
apache/2.4.18          & 0.38      & 0.06   & 0.10              & 922     \\
apache/2.4.39          & 0.62      & 0.43   & 0.51              & 152     \\
nginx/1.14.2           & 0.62      & 0.25   & 0.35              & 840     \\
nginx/1.9.5            & 0.83      & 0.71   & 0.77              & 7       \\
nginx/1.4.4            & 1.00      & 0.07   & 0.13              & 14      \\
openresty/1.19.9.1     & 0.83      & 0.88   & 0.85              & 190     \\
nginx/1.17.0           & 0.43      & 0.43   & 0.43              & 7       \\
nginx/1.17.6           & 0.75      & 0.32   & 0.44              & 38      \\
apache/2.4.49          & 0.00      & 0.00   & 0.00              & 9       \\
apache/2.4.38          & 0.29      & 0.16   & 0.21              & 1049    \\
nginx/1.17.1           & 0.00      & 0.00   & 0.00              & 3       \\
nginx/1.15.6           & 0.88      & 0.39   & 0.55              & 38      \\
nginx/1.2.9            & 0.50      & 0.20   & 0.29              & 5       \\
nginx/1.11.9           & 0.88      & 0.33   & 0.48              & 21      \\
apache/2.2.32          & 0.00      & 0.00   & 0.00              & 10      \\
nginx/1.19.6           & 0.67      & 0.03   & 0.06              & 66      \\
openresty/1.15.8.1     & 0.75      & 0.25   & 0.38              & 24      \\
apache/2.4.51          & 0.90      & 0.22   & 0.35              & 293     \\
nginx/1.21.0           & 0.36      & 0.12   & 0.19              & 40      \\
nginx/1.11.6           & 0.00      & 0.00   & 0.00              & 3       \\
nginx/1.23.2           & 0.00      & 0.00   & 0.00              & 16      \\
apache/2.2.22          & 0.38      & 0.23   & 0.28              & 259     \\
openresty/1.21.4.1     & 0.91      & 0.90   & 0.91              & 187     \\
apache/2.4.12          & 0.00      & 0.00   & 0.00              & 17      \\
nginx/1.9.9            & 0.00      & 0.00   & 0.00              & 7       \\
nginx/1.4.6            & 0.81      & 0.87   & 0.84              & 135     \\
nginx/1.15.10          & 0.00      & 0.00   & 0.00              & 11      \\
nginx/1.13.1           & 0.00      & 0.00   & 0.00              & 5       \\
nginx/1.6.2            & 0.43      & 0.15   & 0.22              & 109     \\
apache/2.2.29          & 0.22      & 0.11   & 0.15              & 36      \\
nginx/1.13.12          & 0.25      & 0.05   & 0.08              & 21      \\
apache/2.4.28          & 0.44      & 0.57   & 0.50              & 7       \\
nginx/1.19.7           & 0.33      & 0.12   & 0.18              & 8       \\
nginx/1.5.0            & 0.00      & 0.00   & 0.00              & 3       \\
nginx/1.21.6           & 0.63      & 0.62   & 0.62              & 545     \\
nginx/1.10.1           & 0.54      & 0.19   & 0.28              & 69      \\
apache/2.4.20          & 0.67      & 0.17   & 0.27              & 12      \\
nginx/1.22.0           & 0.60      & 0.41   & 0.49              & 823     \\
apache/2.4.33          & 0.92      & 0.21   & 0.34              & 52      \\
nginx/1.15.12          & 0.50      & 0.03   & 0.06              & 33      \\
microsoft-iis/10.0     & 0.93      & 0.95   & 0.94              & 5067    \\
nginx/1.16.0           & 1.00      & 0.40   & 0.57              & 5       \\
nginx/1.17.2           & 0.83      & 0.19   & 0.31              & 211     \\
nginx/1.21.4           & 0.00      & 0.00   & 0.00              & 6       \\
apache/2.4.54          & 0.25      & 0.05   & 0.09              & 39      \\
apache/2.4.48          & 0.52      & 0.65   & 0.58              & 2004    \\
nginx/1.8.0            & 0.38      & 0.04   & 0.07              & 238     \\
nginx/1.17.5           & 0.12      & 0.03   & 0.05              & 32      \\
apache/2.3.6           & 0.00      & 0.00   & 0.00              & 30      \\
microsoft-iis/5.1      & 0.31      & 0.52   & 0.39              & 99      \\
nginx/1.15.5           & 0.75      & 0.48   & 0.59              & 25      \\
nginx/1.11.2           & 0.33      & 0.04   & 0.08              & 23      \\
apache/2.2.26          & 0.16      & 0.07   & 0.09              & 417     \\
apache/5.2             & 0.33      & 0.30   & 0.32              & 10      \\
apache/1.3.27          & 0.73      & 0.40   & 0.52              & 20      \\
nginx/1.19.10          & 0.50      & 0.55   & 0.52              & 22      \\
apache/2.4.25          & 0.50      & 0.03   & 0.06              & 29      \\
openresty/1.19.3.1     & 0.26      & 0.13   & 0.17              & 793     \\
nginx/1.4.3            & 0.73      & 0.50   & 0.59              & 16      \\
nginx/1.0.10           & 0.79      & 0.94   & 0.86              & 2164    \\
apache/2.2.10          & 0.97      & 0.99   & 0.98              & 285     \\
microsoft-iis/5.5      & 0.00      & 0.00   & 0.00              & 18      \\
apache/2.4.37          & 1.00      & 0.17   & 0.29              & 6       \\
apache/1.6.0           & 0.25      & 0.34   & 0.29              & 400     \\
nginx/1.16.1           & 0.40      & 0.67   & 0.50              & 6       \\
nginx/1.7.11.1         & 0.37      & 0.29   & 0.33              & 1101    \\
nginx/1.13.8           & 0.14      & 0.25   & 0.18              & 4       \\
nginx/0.8.54           & 0.00      & 0.00   & 0.00              & 4       \\
nginx/1.13.11          & 0.32      & 0.26   & 0.29              & 35      \\
nginx/1.7.6            & 0.92      & 0.92   & 0.92              & 3206    \\
nginx/1.19.4           & 0.79      & 0.17   & 0.27              & 66      \\
nginx/1.7.0            & 0.86      & 0.58   & 0.69              & 31      \\
nginx/1.18.0           & 0.00      & 0.00   & 0.00              & 6       \\
nginx/0.5.32           & 0.43      & 0.82   & 0.56              & 4193    \\
apache/2.5.1           & 0.00      & 0.00   & 0.00              & 3       \\
openresty/1.9.7.3      & 0.67      & 0.67   & 0.67              & 3       \\
nginx/1.9.13           & 1.00      & 0.33   & 0.50              & 3       \\
apache/2.4.41          & 1.00      & 0.20   & 0.33              & 5       \\
nginx/1.9.0            & 0.28      & 0.48   & 0.35              & 1817    \\
nginx/1.17.10          & 0.21      & 0.20   & 0.21              & 15      \\
nginx/1.7.2            & 0.94      & 0.58   & 0.72              & 109     \\
apache/2.4.29          & 0.00      & 0.00   & 0.00              & 9       \\
apache/2.2.15          & 0.24      & 0.36   & 0.28              & 1604    \\
apache/2.0.58          & 0.69      & 0.37   & 0.48              & 638     \\
apache/2.2.27          & 0.00      & 0.00   & 0.00              & 8       \\
apache/2.4.43          & 0.22      & 0.08   & 0.12              & 24      \\
openresty/1.5.8.1      & 0.24      & 0.05   & 0.08              & 78      \\
apache/2.4.53          & 0.00      & 0.00   & 0.00              & 3       \\
microsoft-iis/1.18.0   & 0.33      & 0.22   & 0.26              & 384     \\
nginx/1.10.0           & 0.64      & 0.46   & 0.54              & 63      \\
apache/2.4.3           & 0.67      & 0.08   & 0.14              & 25      \\
nginx/1.0.8            & 0.32      & 0.13   & 0.19              & 428     \\
nginx/1.19.8           & 0.00      & 0.00   & 0.00              & 5       \\
nginx/1.0.0            & 0.57      & 0.13   & 0.21              & 31      \\ \midrule
\textbf{micro avg}     & 0.58      & 0.58   & \textbf{0.58}     & 45718   \\
\textbf{macro avg}     & 0.24      & 0.16   & \textbf{0.17}     & 45718   \\
\textbf{weighted avg}  & 0.59      & 0.58   & \textbf{0.55}     & 45718   \\ \bottomrule
\end{longtable}
\clearpage
\twocolumn

    \subsection{Feature importances}\label{sec:featureimportances}

\begin{table*}
\caption{List of test cases, sorted by Gini impurity, as reported by a Random Forest classifier.}
\label{tab:feature-importances}
\begin{tabular}{@{}ccc@{}}
    \toprule
\textbf{Test case} & \textbf{Protocol} & \textbf{Gini impurity} \\ \midrule
7                  & HTTP/80           & 0.0057                 \\
2                  & HTTP/80           & 0.0018                 \\
7                  & HTTPS/443         & 0.0015                 \\
9                  & HTTPS/443         & 0.0010                 \\
9                  & HTTP/80           & 0.0007                 \\
4                  & HTTPS/443         & 0.0006                 \\
8                  & HTTPS/443         & 0.0005                 \\
5                  & HTTP/80           & 0.0005                 \\
4                  & HTTP/80           & 0.0004                 \\
2                  & HTTPS/443         & 0.0003                 \\
6                  & HTTPS/443         & 0.0003                 \\
3                  & HTTPS/443         & 0.0003                 \\
8                  & HTTP/80           & 0.0003                 \\
11                 & HTTPS/443         & 0.0002                 \\
14                 & HTTPS/443         & 0.0002                 \\
13                 & HTTPS/443         & 0.0002                 \\
1                  & HTTPS/443         & 0.0001                 \\
11                 & HTTP/80           & 0.0001                 \\
6                  & HTTP/80           & 0.0001                 \\
12                 & HTTPS/443         & 0.0001                 \\
3                  & HTTP/80           & 0.0001                 \\
10                 & HTTP/80           & 0.0001                 \\
5                  & HTTPS/443         & 0.0001                 \\
10                 & HTTPS/443         & 0.0001                 \\
1                  & HTTP/80           & 0.0000                 \\
14                 & HTTP/80           & 0.0000                 \\
15                 & HTTP/80           & 0.0000                 \\
15                 & HTTPS/443         & 0.0000                 \\
13                 & HTTP/80           & 0.0000                 \\
12                 & HTTP/80           & 0.0000                 \\
0                  & HTTP/80           & 0.0000                 \\
0                  & HTTPS/443         & 0.0000                 \\ \bottomrule
\end{tabular}
\end{table*}

\end{document}